\documentclass{article}

\usepackage{arxiv}

\usepackage[utf8]{inputenc} 
\usepackage[T1]{fontenc}    
\usepackage{hyperref}       
\usepackage{url}            
\usepackage{booktabs}       
\usepackage{amsfonts}       
\usepackage{nicefrac}       
\usepackage{microtype}      
\usepackage{lipsum}
\usepackage{amsmath,amssymb}
\usepackage{graphicx}

\title{Measles-induced immune amnesia and its effects in concurrent epidemics}

\author{
  Guillermo B. Morales\\
 Departamento de Electromagnetismo y F{\'i}sica de la Materia\\
  e Instituto Carlos I de F{\'i}sica Te{\'o}rica y
  Computacional. \\
  Universidad de Granada\\
   E-18071 Granada, Spain\\
  \texttt{guillermobm@onsager.ugr.es} \\
   \And
 Miguel A. Mu{\~n}oz\\
 Departamento de Electromagnetismo y F{\'i}sica de la Materia\\
  e Instituto Carlos I de F{\'i}sica Te{\'o}rica y
  Computacional. \\
  Universidad de Granada\\
   E-18071 Granada, Spain\\
  \texttt{mamunoz@onsager.ugr.com} \\
}

\begin{document}
\maketitle

\begin{abstract}
It has been recently discovered that the measles virus can wipe out the adaptive immune system, destroying B lymphocytes and reducing the diversity of non-specific B cells of the infected host.  In particular, this implies that previously acquired immunization from vaccination or direct exposition to other pathogens could be erased in a  phenomenon named "immune amnesia", whose effects can become particularly worrisome given the actual rise of anti-vaccination movements. Here we present the first attempt to incorporate immune amnesia into standard models of epidemic spreading. In particular, we analyze diverse variants of a model that describes the spreading of two concurrent pathogens causing measles and another generic disease:  the SIR-IA model. Analytical and computational studies confirm that immune amnesia can indeed have important consequences for epidemic spreading, significantly altering the vaccination coverage required to reach herd-immunity for concurring infectious diseases.  More specifically, we uncover the existence of novel propagating and endemic phases which are induced by immune amnesia, that appear both in fully-connected and more structured networks, such as random networks and power-law degree-distributed ones. In particular, the transitions from a quiescent state into these novel phases can become rather abrupt in some cases that we specifically analyze. Furthermore, we discuss the meaning and consequences of our results and their relation with, e.g., immunization strategies, together with the possibility that explosive types of transitions may emerge, making immune-amnesia effects particularly dramatic. This work opens the door to further developments and analyses of immune amnesia effects, contributing, more generally, to the theory of interacting epidemics on complex networks.
\end{abstract}


\section{Introduction}
The measles virus is among the most contagious human pathogens; it
can cause severe symptoms and death, mostly during childhood and, as such,
it represents a serious problem for global public health, targeted by the World
Health Organization
(WHO) \cite{noauthor_who_nodate, MV}. 
In spite of the $73\%$ global drop in measles
deaths achieved thanks to improved vaccination policies in the period
$2000-2018$, measles is still common in many developing countries;
indeed, over $500.000$ cases were reported worldwide in $2019$, more
than a half of which in Africa \cite{noauthor_who_nodate-1}.  Also in
the USA as well as in Europe, where measles is considered endemic in at least ten countries, outbreaks are becoming ubiquitous in recent years ---with a $30\%$ overall increase from 2017 to 2018---  mostly as a consequence of anti-vaccination movements \cite{Eva,Italy}.
Moreover, the WHO has recently raised the alarm
over the increasing chance of measles outbreaks due to poor
vaccination coverage as the Covid-19 pandemic progresses, with millions of  children at risk of missing out on measles vaccines \cite{noauthor_who_nodate-2}.

Given the magnitude of the problem, it should come as no surprise
that, as of $2016$, there were already over $100$ mathematical models
 proposed in the literature to specifically reproduce and predict the
evolution of measles outbreaks \cite{thompson_evolution_2016}. In
spite of this wealth of modeling approaches, there is a crucial aspect of
measles that is still systematically neglected and that has the potentiality to be more harmful than the outbreaks
themselves: \emph{immune amnesia}.

Building over a previous series of works that linked childhood
mortality and severe immunosuppression with preceding measles-virus
infection
\cite{de_vries_measles_2012,de_vries_measles_2014,mina_long-term_2015}, conclusive empirical evidence has been very recently found
that measles can wipe out acquired immunity to other infectious diseases
through a mechanism called ``\emph{immune amnesia}''
\cite{mina_measles_2019,petrova_incomplete_2019,IA-Netherlands}. More specifically, measles infection has been shown to destroy B lymphocytes (specific to whichever other pathogens) and to reduce the diversity of non-specific B cells, thus limiting severely the acquired defenses in the adaptive immune system (regardless of whether these have been achieved by means of vaccination or direct contact with a pathogen) \cite{mina_measles_2019,petrova_incomplete_2019,IA-Netherlands,Eva}. In fact, previous studies with rhesus macaques \cite{de_vries_measles_2012}, as well as with unvaccinated children \cite{mina_measles_2019,IA-Netherlands}, had measured a depletion of
up to $70\%$ of the existing antibody repertoire across individuals
after measles infection, even if there is a large subject-to-subject
variability. As a matter of fact, it has been long reported that the majority  of measles-related deaths are not due to the measles virus itself, but to secondary infections caused by the associated immunosuppression \cite{miller_frequency_1964,beckford_factors_1985}, hence the importance of taking into account immune amnesia into the broader field of
epidemic mathematical modelling \cite{anderson_infectious_1992}.

In this work, we give a first step toward bridging this gap by incorporating the possibility of measles-induced immune amnesia onto standard models of epidemic spreading for an arbitrary infectious disease co-occurring  with measles outbreaks. Starting from an initial situation where vaccination coverage is assumed to grant herd immunity for a certain infectious disease ---i.e. a sufficient number of vaccinated individuals so that the disease can hardly spread across the population---, could measles outbreaks wipe out
such a immunity to the point where sizeable secondary epidemics are unleashed?
If that was the case, the aforementioned recent increase in measles
outbreaks worldwide could be a greater threat than previously thought.
Even worse,  a potential herd-immunity
strategy relying  on  vaccination for COVID-19 could be hindered by
the effects of measles outbreaks, all the most in countries where measles vaccination coverage during this health crisis is at its lowest.

To analyze these issues, here we develop a simple mathematical model that sheds light on  the  effects of immune amnesia over the dynamics of a second epidemic disease co-existing with outbreaks of measles. In particular, we perform mathematical analyses and extensive computer simulations of a modified Susceptible-Infectious-Recovered (SIR) model  that accounts for two coupled diseases, vaccination coverage, and possible demographic effects, as well as, crucially, the possibility of immune amnesia. We start considering homogeneously-mixed populations, i.e. fully connected networks, and perform standard mean-field calculations that allow us to derive, e.g., analytical estimates for the minimum measles-vaccination coverage needed to maintain herd-immunity for the second epidemics. Then, we extensively  analyze, both theoretically and computationally, the impact that the structure of the underlying network of contacts can have on the results. In all cases, we elucidate the possible emergence of immune-amnesia induced phases, where the X-disease becomes propagating/endemic just as a consequence of immune-amnesia effects.

\section{The SIR-IA model}

In order to mimic the effects of immune amnesia on a given population, we extend the SIR model ---either with or without demographic dynamics--- \cite{Hethcote,Anderson-book,Murray-book,Pastor-Satorras} 
to account for two co-occurring diseases: measles (M) and a second generic infectious disease to which  we refer by X hereon. For the sake
of illustration, we consider X to be COVID-19 as a guiding example, and use its associated epidemic parameters. We name this SIR-like model with Immune Amnesia (IA), SIR-IA model. 

\begin{figure}
\begin{centering}
\includegraphics[scale=0.45]{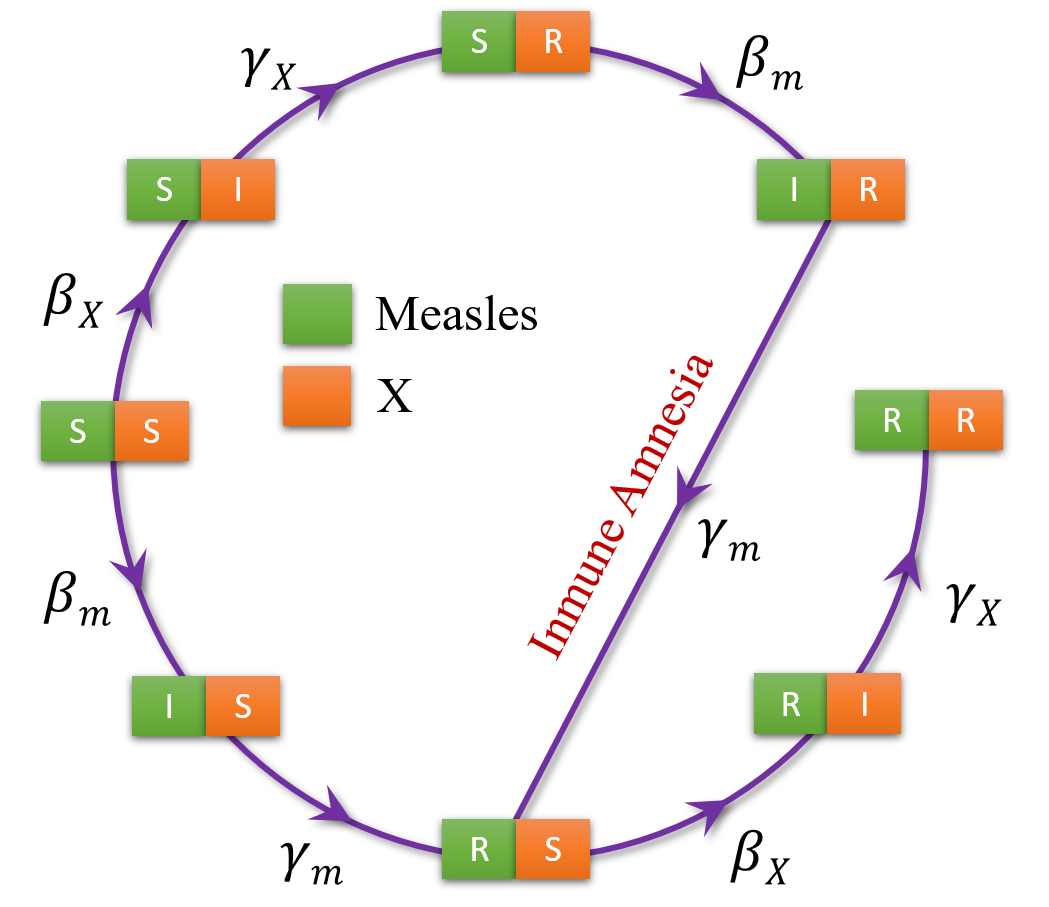}
\par\end{centering}
\caption{\textbf{Sketch of the transitions between the eight allowed  system states.} Green (orange) cells correspond to measles (X) disease states. For the sake of clarity, demographic processes are not included.\label{fig:Model_Scheme}}
\end{figure}

As in the standard SIR dynamics \cite{Hethcote,Anderson-book,Murray-book,Pastor-Satorras},  in the SIR-IA model each of the N individuals within the focus population can be either susceptible to be infected (S), infected (I), or resistant (R),  for each of the two diseases. Thus, there are a total of nine possible states: $i,j \in \{SS,SI,SR,IS,II,IR,RS,RI,RR\}$, denoting  the state of individuals that are simultaneously in state $i\in\left\{ S,I,R\right\}$ for measles and $j\in\left\{ S,I,R\right\}$ for disease X.  It is important to remark that: 
\begin{itemize}
\item States $II$ representing individuals infected simultaneously of both diseases  will be dismissed  in first approximation as highly unlikely, given the short recovery periods.
\item Resistant populations include not only recovered individuals, but also those who achieved immunity through vaccination.
\end{itemize}
The model dynamics is defined by a Master equation including the set of possible transition rates between these states. This can be numerically integrated in an exact way by using the Gillespie algorithm \cite{Gillespie} (see below).
The set of possible transitions between states,  together with the corresponding rates at which they occur, are schematically depicted in Fig.\ref{fig:Model_Scheme} (see also Table \ref{table1} for a definition of model parameters together with their base-line values). 
In particular, following the standard notation, $\beta_{M},\,\beta_{X}$ and $\gamma_{M},\,\gamma_{X}$ denote the infectivity and recovery rates for measles and disease X, respectively. Notice that IA is explicitly implemented within the term $+\gamma_{M}\rho_{IR}$ in Eq.(\ref{eq:5}), which drags X-recovered individuals  back into the pool of X-susceptible ones, in the case they had been infected with measles. The parameters, $v_{X}$ and $v_{M}$ represent the vaccination coverage for each disease, i.e., the probability that a new individual added into the system is vaccinated of measles and disease X, respectively. For COVID-19, it has been estimated that around a 65\% of the population should be resistant (either through vaccination or naturally acquired immunity) in order to reach herd-immunity \cite{fontanet_covid-19_2020}. Hence, to study to what extent IA effects can affect a potentially achieved COVID-19 herd-immunity, we consider a hypothetical large vaccination coverage value, $v_{X}=0.9$.

\begin{table}
\centering
\caption{Epidemic parameters used to model measles (M) and COVID-19 (X) epidemics. Infectivity and recovery periods for each disease were taken from \cite{bauch_transients_2003,guerra_basic_2017,li_substantial_2020} to match the reported $R_{0}$ values (e.g., for COVID-19 we took 
$R_{0} = 3.4$; although recent studies suggest a shorter infectivity period for COVID-19, this does not significantly affect the forthcoming conclusions). COVID-19 vaccination coverage was set to a high-value to ensure initial herd-immunity in the homogeneous-mixing approach.\label{table1}}
\begin{tabular}{lrrr}
& Symbol & Base-line value \\
\midrule
Measles infectivity rate & $\beta_{M}$ & $2.1$~days$^{-1}$ \\
Measles recovery rate & $\gamma_{M}$ & $1/8$~days$^{-1}$ \\
Covid-19 infectivity rate & $\beta_{X}$ & $0.25$~days$^{-1}$\\
Covid-19 recovery rate & $\gamma_{X}$ & $1/14$~days$^{-1}$\\
Immigration/emigration rate & $\mu$ & 1/365 days$^{-1}$\\
Covid-19 vaccination coverage & $v_{X}$ & 0.9\\
\bottomrule
\end{tabular}


\end{table}
As for "demographic" parameters, the death and birth rates for all individuals ---regardless of their possible disease state--- have been set to a common value $\mu$. These rates can also be interpreted ---looking at the problem from  a meta-population perspective---  as describing emigration and immigration processes. In particular, this latter interpretation justifies the use of relatively large rate values (see Table \ref{table1}). In individual-based stochastic simulations of the model, any removed individual is instantaneously replaced by a new-arrived one, thus keeping a fixed population size. 

For the sake of simplicity, we begin by studying the  case 
of homogeneously-mixed populations and then analyze more structured 
populations with a non-trivial underlying network of contacts.
We study versions of the model
with either no explicit demography (i.e. $\mu=0$) or explicit demographic effects $\mu\neq0$. In the first case, much as in the standard SIR model, there cannot possibly be any non-trivial stationary endemic state, while in the second such states can possibly exist \cite{anderson_infectious_1992}. We investigate in parallel all these  possible scenarios to illustrate the generality of the conclusions from complementary perspectives.

\section{Results}

\subsection{Homogeneously-mixed populations}

To gain insight into the model key features, we employ a standard mean-field approximation which, as usual,  is exact in the limit of infinitely large homogeneously-mixed populations. This, leads rather straightforwardly to  the following set of eight differential equations (sometimes called "rate equations") \cite{Hethcote}:

\begin{subequations}
\begin{eqnarray}
&\dot{\rho}_{SS}=& -\beta_{X}\rho_{SS}(\rho_{SI}+\rho_{RI})-\beta_{M}\rho_{SS}(\rho_{IS}+\rho_{IR})\label{eq:1}
\\&&-\mu\rho_{SS}+\mu(1-v_{X}(1-v_{M})-v_{M}(1-v_{X})-v_{M}v_{X}) \nonumber,\\
&\dot{\rho}_{SI}= & \beta_{X}\rho_{SS}\left(\rho_{SI}+\rho_{RI}\right)-\gamma_{X}\rho_{SI}-\mu\rho_{SI},\label{eq:2}\\
&\dot{\rho}_{SR}= & -\beta_{M}\rho_{SR}\left(\rho_{IS}+\rho_{IR}\right)+\gamma_{X}\rho_{SI}-\mu\rho_{SR}\label{eq:3}
\\&&+\mu v_{X}(1-v_{M}) \nonumber,\\
&\dot{\rho}_{IS}= & \beta_{M}\rho_{SS}\left(\rho_{IS}+\rho_{IR}\right)-\gamma_{M}\rho_{IS}-\mu \rho_{IS},\label{eq:4}\\
&\dot{\rho}_{RS}= & -\beta_{X}\rho_{RS}\left(\rho_{SI}+\rho_{RI}\right)+\gamma_{M}\left(\rho_{IS}+\rho_{IR}\right)\label{eq:5}
\\&&-\mu\rho_{RS}+\mu v_{M}(1-v_{X}) \nonumber,\\
&\dot{\rho}_{IR}= & \beta_{M}\rho_{SR}\left(\rho_{IS}+\rho_{IR}\right)-\gamma_{M}\rho_{IR}-\mu\rho_{IR},\label{eq:6}\\
&\dot{\rho}_{RI}= & \beta_{X}\rho_{RS}\left(\rho_{SI}+\rho_{RI}\right)-\gamma_{X}\rho_{RI}-\mu\rho_{RI},\label{eq:7}\\
&\dot{\rho}_{RR}= & \gamma_{X}\rho_{RI}-\mu\rho_{RR}+\mu v_{M}v_{X}\label{eq:8}
\end{eqnarray}
\end{subequations}

\noindent where $\rho_{ij}$ is the population fraction in state $ij$. 

\subsubsection{SIR-IA model without demography}
Let us begin by analyzing the case with no demography, i.e. with $\mu=0$. In this scenario, the only role of $v_M$ and $v_X$ is to determine the initial fraction of vaccinated population of M and X diseases, respectively. To simplify the forthcoming mathematical analyses, let us define $i_{M}=\rho_{IS}+\rho_{IR}$ and $s_{M}=\rho_{SS}+\rho_{SR}$ as the total fraction of infectious and susceptible individuals of measles, and their counterparts $i_{X}=\rho_{SI}+\rho_{RI}$ and $s_{X}=\rho_{SS}+\rho_{RS}$ for the disease X. Let us remark that states $SI$ and $IS$ are not counted as susceptible ones (neither for measles nor for X disease) since states $II$  have been neglected, as previously said.

To analyze the situation in which herd immunity for the disease X is potentially erased by the effect of immune amnesia, we assume that the population fraction $v_{X}$ vaccinated of X is sufficiently high to initially avoid the spreading of the disease (i.e. $i_{X}\approx0$ prior to the onset of a measles outbreak). Notice that, under this assumption,  from Eqs.(\ref{eq:1}),(\ref{eq:3}),(\ref{eq:5}) and (\ref{eq:6}) one readily derives the standard SIR-model mean-field equations for measles dynamics:
\begin{eqnarray}
&&\dfrac{di_{M}}{dt} =  \beta_{M}i_{M}s_{M}-\gamma_{M}i_{M}\label{dim_dt}\\
&&\dfrac{ds_{M}}{dt}  =  -\beta_{M}i_{M}s_{M}\label{dsm_dt}.
\end{eqnarray}
From Eq.(\ref{dim_dt}) it is clear that a small  seed (of size $\epsilon\ll N$) of M-infectious individuals  introduced in a population of initially susceptible and resistant individuals grows exponentially if $\frac{di_{M}}{dt}>0$, and decays to $0$ if this is negative.  Therefore, since {$s_{M}(0)=(N (1-v_M)-\epsilon)/N\approx 1-v_M$} is the initial fraction of  M-susceptible to measles, the epidemic threshold separating the above two regimes is specified by the condition: 
\begin{equation}
v_{M}^{\dagger}=1-1/R_{0}^{M} \label{Rm}
\end{equation}
where, as usual, the basic reproduction number  $R_{0}^{M}=\frac{\beta_{M}}{\mu+\gamma_{M}}$ {---defined as  the  average number of secondary infections generated by a primary case  in  a  completely  susceptible  population \cite{anderson_infectious_1992}---} has been introduced for measles. Therefore, $v_{M}^{\dagger}$ is the herd-immunity threshold and represents the minimum fraction of M-vaccinated population needed to prevent measles spreading.  For  the considered parameter values (see Table \ref{table1}) it follows that $R_{0}^{M}\approx 17$ and $v_M^{\dagger}\approx0.95$, numbers that emphasize
the well-known high infective power of the measles virus. 

In what respects the X disease and \emph{in the absence of IA}, one can easily derive from Eqs.(\ref{eq:1}),(\ref{eq:2}),(\ref{eq:4}) and (\ref{eq:7}) an epidemic-threshold condition analogous to Eq.(\ref{Rm}):
\begin{equation}
v_{X}^{\dagger}=1-1/R_{0}^{X} 
\label{Rx},
\end{equation}
which results into $R_{0}^{X}\approx 3.5$ and $v_X^{\dagger}\approx0.65$ for the parameters in Table \ref{table1}. Let us underline that, in this IA-free case, both diseases are uncoupled and, hence, their respective thresholds are  independent of each other.
\begin{figure}
\begin{centering}
\includegraphics{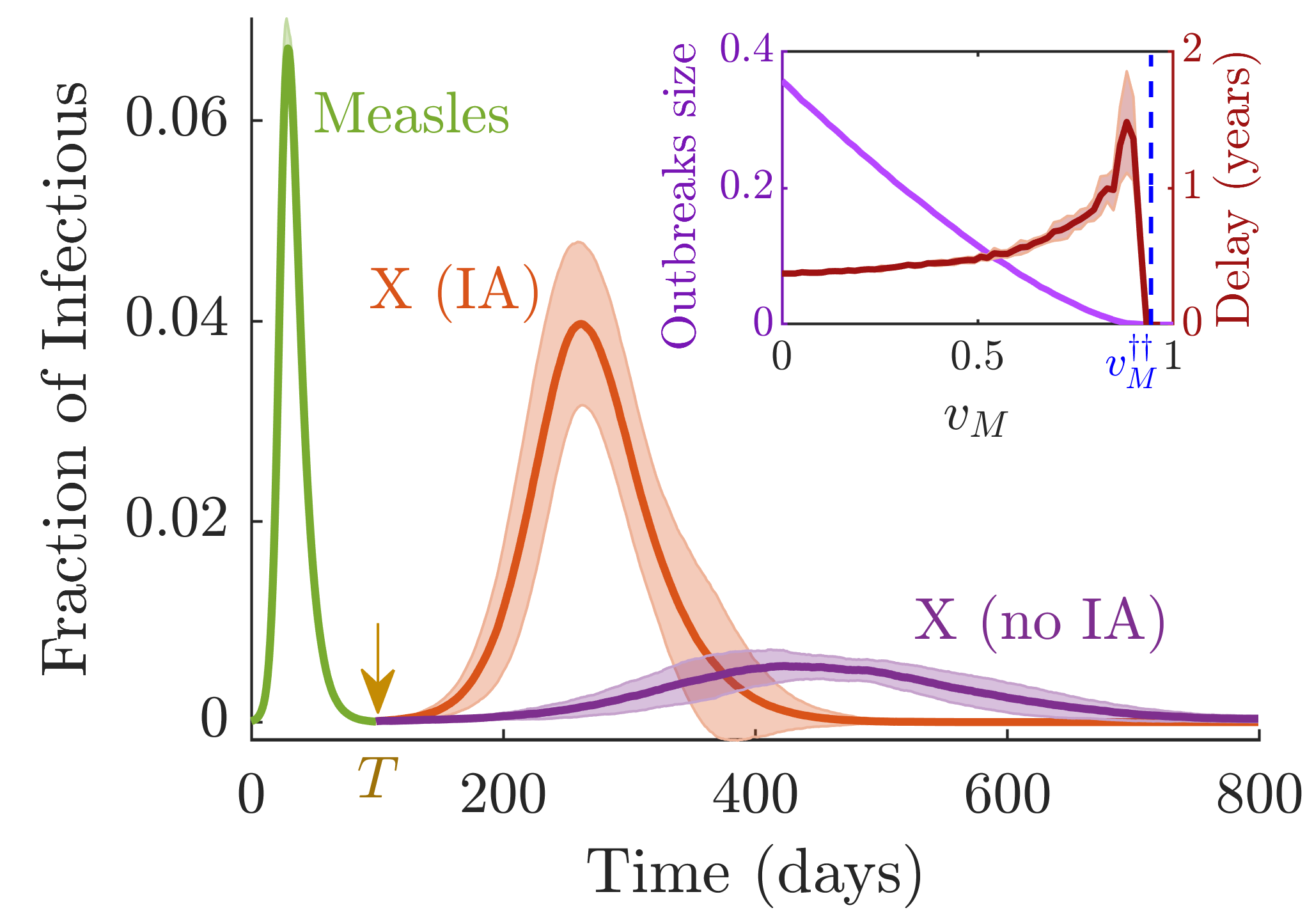}
\par\end{centering}
\caption{\textbf{Time evolution of the  fraction of:  (i) M-infectious individuals
(green curve), (ii) X-infectious individuals in the absence of IA (purple), and (iii) X-infectious individuals under the presence of IA effects (orange).} X-vaccination coverage was set slightly over its herd-immunity threshold $v_{X}=v_X^\dagger(1+\varepsilon)$, with $\varepsilon=0.001$. Conversely, $v_{M}=0.8<v_M^\dagger$ was set to guarantee the spreading of measles outbreaks. At time $T=100 days$, a seed $\epsilon=10^{-4}N$ of X-infectious individuals was inserted into the system. In the \textbf{inset}, we plot the maximum size of the X outbreak (measured as the total fraction of infectious individuals integrated in time) and the time elapsed between the green and orange peaks (in years) against M-vaccination coverage. The blue-dashed line marks the theoretical approximation to the vaccination threshold (see Eqs.(30) and (31) in the SI). Simulations were performed within the mean-field, homogeneous-mixing approximations, using the Gillespie algorithm with a total population size $N=10^{5}$. Shaded areas around the curves indicate one standard deviation, as determined from over $50$ independent runs.  Parameters were chosen according to Table \ref{table1}.} \label{fig:SIR_Evolution}
\end{figure}

\begin{figure}
\begin{centering}
\includegraphics{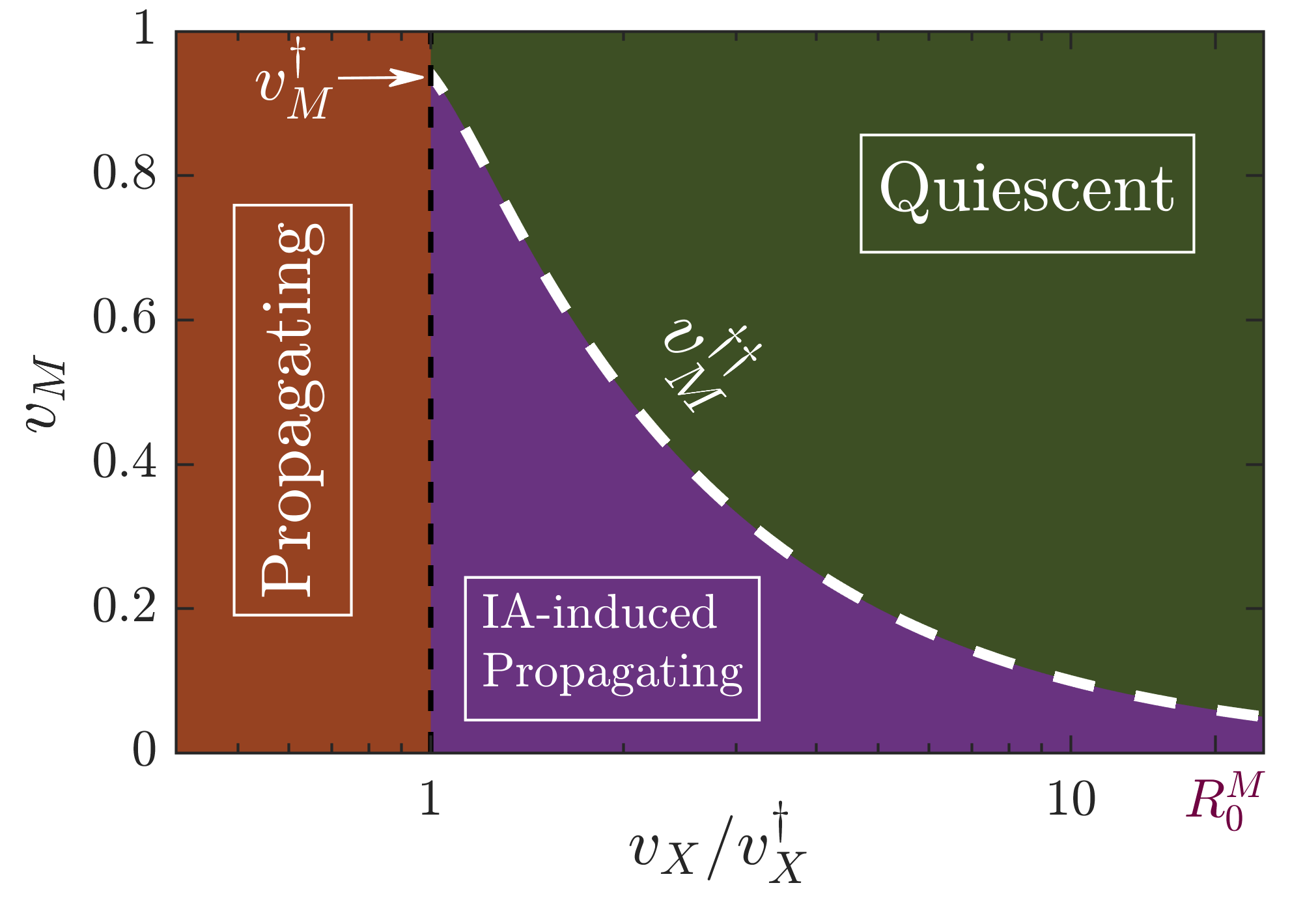}
\par\end{centering}
\caption{\textbf{Analytically determined phases for the X disease in the SIR-IA model without demography}. The white-dashed line depicting the vaccination threshold under IA effects was determined by solving Eqs.(30) and (31) in Appendix B.1. from the SI. A lower bound for the IA-induced endemic phase is given by $v_{X}=v_{X}^{\dagger}$, below which the disease is always endemic, independently of IA effects (black-dashed line). Notice that a triple point appears for $v_X=v_X^\dagger$ and $v_M=v_M^\dagger$ (marked by the white arrow), at which the three phases collide. 
\label{fig:Phase_Space}}
\end{figure}
To start scrutinizing the full problem, including immune-amnesia ---which turns out to be much more intricate from an analytical viewpoint--- we start by performing computational analyses of the stochastic model (see Methods for technical details). In particular, we run simulations of the  SIR-IA  stochastic model  by implementing a Gillespie algorithm as follows: beginning with an initial seed of $\epsilon_{M}$  M-infectious individuals, we let an outbreak of measles spread through the population (for which we set $v_{M}<v_{M}^\dagger$); once it fades out, we add at some initial time $T$ a second seed consisting  of $\epsilon_{X}$ X-infectious individuals who will potentially spread disease X through the system. Fig.\ref{fig:SIR_Evolution} illustrates the 
resulting time courses of epidemics as obtained from
stochastic simulations averaged over many realizations. In particular, for the case in which the vaccination coverage $v_{X}$ is only slightly above the herd-immunity threshold $v_{X}^\dagger$ (i.e. $v_{X}\gtrsim v_{X}^\dagger$) the figure clearly shows that ---on average---  much larger outbreaks occur under the influence of IA. It also reveals that, not surprisingly, the duration of the outbreaks is shorter when IA-effects are considered, as the disease takes over the susceptible population in a much-faster way. 
The inset of Fig.\ref{fig:SIR_Evolution} illustrates how 
these  results change quantitatively with the M-vaccination coverage, $v_M$: as naively expected, (i) the total outbreak size grows and (ii) the time elapsed between measles and X outbreak peaks becomes shorter as $v_M$ is reduced. Observe also that X epidemics can only break out (i.e. become supercritical) if $v_M$ is set below a minimum, \emph{critical vaccination threshold}  $v_M^{\dagger\dagger}(\mu=0)<v_M^{\dagger}$, which represents \emph{the minimum population fraction that needs to be vaccinated of measles in order to preserve herd-immunity for disease X even in the presence of IA effects}. 

Last but not least, let us stress that, in spite of the fact that one needs to deal with a set of $8$ differential equations, we have been able to find an analytical solution for $v_M^{\dagger\dagger}(\mu=0)$ as a function of other parameters. The detailed calculations, which are obtained as a limiting case of the more general problem of a general network architecture, can be found in Appendix B.1. Fig.\ref{fig:SIR_Evolution} confirms that the analytically-derived threshold (blue dotted line in the inset), is in excellent  agreement with computational results, a conclusion that remains true for other choices of parameter values. A summary of the analytical and computational results
is provided by Fig.\ref{fig:Phase_Space} which explicitly illustrates
the existence, for a broad range of parameter values, of a propagating phase which emerges  as a mere consequence of immune-amnesia effects.

\begin{figure}
\begin{centering}
\includegraphics{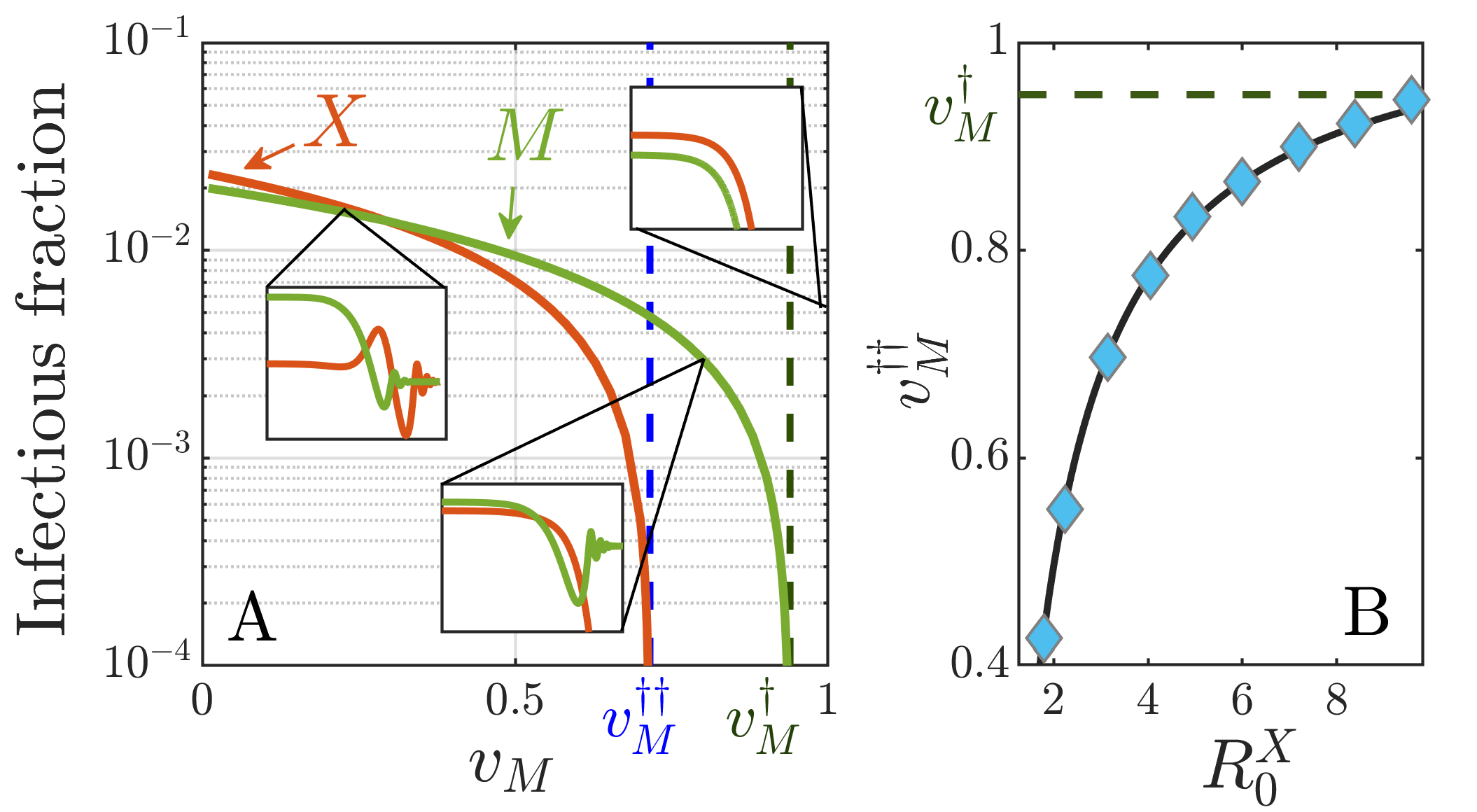}
\par\end{centering}
\caption{\textbf{Analysis of the stationary states and epidemic thresholds in the SIR-IA with demography.} \textbf{A}: stable stationary state for the X-infectious (orange line) and M-infectious (purple line) population fractions as a function of the measles vaccination coverage $v_{M}$ (vanishing values in the disease-free steady states are not shown). Blue and green dashed lines mark the analytically-found critical points $v_{M}^{\dagger\dagger}=0.72$ and $v_{M}^{\dagger}=0.94$, respectively, according to Eq.(\ref{crit_X}). In the \textbf{insets}, we illustrate the time evolution (in days) of the X-infectious (orange) and M-infectious (green) population fractions for the three different regimes found in Fig.\ref{fig:Phase_Space2}. Curves where obtained by direct integration of  Eqs.(\ref{eq:1}-\ref{eq:8}). \label{fig:Bifurcation-Diagram} \textbf{B}: minimum fraction of M-vaccinated individuals, $v_{M}^{\dagger\dagger}$, necessary to prevent an endemic state of disease X as a function of the reproduction number $R_{0}^{X}$. The black curve represents the theoretical approximation for $v_{M}^{\dagger\dagger}$ according to Eq.(\ref{crit_X}), while blue markers correspond to the numerical solutions obtained by studying the fixed points of the system and their stability (see Methods). All parameters were chosen according to Table \ref{table1}.
\label{fig:Minimum-fraction-of}}
\end{figure}

\subsubsection{SIR-IA model with demography}

In general, introducing demographic dynamics such as birth/death and/or emigration/immigration processes into a SIR-like model, opens up the possibility for stable endemic states, with a non-zero fraction of infectious individuals, to appear. For this, it is necessary  that such processes occur at a fast-enough pace so that a flux of new susceptible individuals is constantly generated to "feed" the contagion process, otherwise the epidemics necessarily vanish
\cite{anderson_infectious_1992}.

As a first test to analyze the demographic version of the model, with $\mu \neq 0$, we verified the existence of endemic states by numerically solving the mean-field Eqs.(\ref{eq:1}-\ref{eq:8}) with $\mu>0$ (see Table I) and X-vaccination coverage $v_{X}>v_{X}^\dagger$, for which the X-disease-free state is stable in the absence of IA effects. In particular, the green curve in Fig.\ref{fig:Bifurcation-Diagram}A shows that a measles endemic state is found as soon as the fraction $v_M$ drops below a certain measles herd-immunity level ${v_{M}^\dagger\approx 0.95}$. This result  has been also verified by means of direct Gillespie simulations of the full stochastic model as well as proved analytically, as shown below. Fig.\ref{fig:Bifurcation-Diagram}A also reveals that ---even if it has been obtained for a large X-vaccination coverage value--- an X-disease endemic state exists (orange curve) provided $v_M$ drops below a certain critical threshold value ${v_M^{\dagger\dagger}(\mu \neq 0)}$. Such an X-disease endemic state is purely induced by IA effects, i.e. it is a \textit{immune-amnesia-induced endemic phase}.
\footnote{Let us remark that $v_M^{\dagger\dagger}$ depends on $\mu$ and, in particular, it does not coincide for the cases with and without demography.}

\begin{figure}
\begin{centering}
\includegraphics{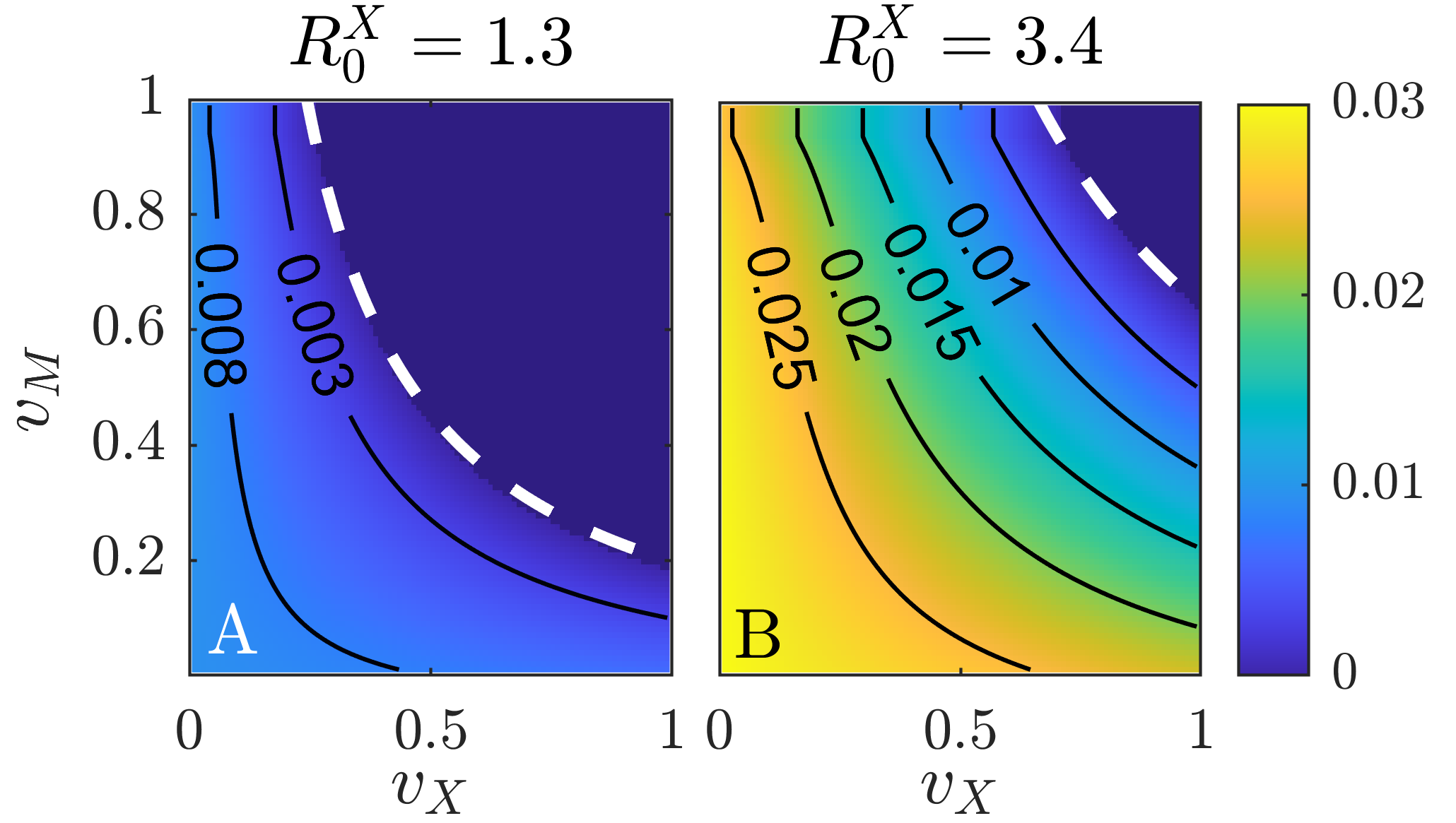}
\par\end{centering}
\caption{\textbf{Fraction of infectious individuals for disease X (color coded) in the
stationary state, against vaccination coverage for both diseases.} White-dashed lines indicate the theoretical approximation to the critical point for each $v_{X}$ value, as given by Eq.(\ref{crit_X}); population size $N=10^5$. \textbf{A}:
results for a mildly infectious disease, with $R_{0}^{X}=1.3$ slightly over SIR critical threshold. \textbf{B}: results for a more virulent disease, with parameters akin to those of COVID-19, as recorded in Table \ref{table1}, i.e. $R_{0}^{X} \approx3.4$ Each point in the grid was obtained numerically by solving for the fixed points of Eqs.(\ref{eq:1}-\ref{eq:8}). \label{fig:HeatMap}}
\end{figure}
We also analyzed the dependence of $v_M^{\dagger\dagger}$ on $R_{0}^{X}$ by solving for the stationary states of Eqs.(\ref{eq:1}-\ref{eq:8}); the endemic states are represented by blue diamonds in Fig.\ref{fig:Minimum-fraction-of}B. Clearly, the larger the virulence of the X-disease the larger the value of 
$v_M^{\dagger\dagger}$. Observe in particular, that as $R_{0}^{X}\rightarrow\infty$,  $v_{M}^{\dagger\dagger}$ approaches that of the critical point for measles outbreak $v_{M}^{\dagger\dagger}\approx v_{M}^{\dagger}$, implying that as soon as an outbreak of measles takes place, an epidemic of $X$ may emerge taking over the newly generated pool of IA-induced susceptible individuals, in spite of the fact that the population was massively vaccinated for the X disease ($v_{X}\gg v_{X}^\dagger$).  Similarly, Fig.\ref{fig:HeatMap} shows the fraction of infected individuals in the stationary state (color coded) as a function of both, $v_M$ and $v_X$. Results for two different X diseases are shown: a mildly infectious one  with  (A) $R_0=1.3$  and (B) COVID-19, with an estimated $R_0=3.4$. It can be observed that, not surprisingly, the region in the vaccination parameter space where X-disease endemic states appear becomes larger with increasing $R_0^X$, but in both cases ---even in the limit of stringent vaccination policies for the X-disease, $v_X \approx 1$--- an IA-induced endemic state can appear if M-vaccination drops below a critical threshold level $v_{M}^{\dagger\dagger}$. The exact value of this threshold (which marks the boundary of the stable endemic region) will depend on $v_X$.
Notice also that the shift between endemic and disease-free states is a gradual one, as corresponds to a continuous phase transition in the present case of homogeneously-mixed populations.

In summary, the value of the transition point $v_{M}^{\dagger\dagger}(\mu \neq 0)$ has been numerically shown to depend on both the infectivity $R_0^X$ and the vaccination coverage $v_X$ for the disease X (as well as on other parameters such as $\mu$). In what follows, in order to get a deeper understanding of the phenomenon as well as a better quantitative 
description of the phase diagram we derive analytical expressions for such dependencies.

\subsubsection{SIR-IA model with demography: theoretical results}
 
Let us first analyze the possible stable fixed points of Eqs.(\ref{eq:1}-\ref{eq:8}) as a function of $v_{M}$. Starting from Eqs. (\ref{eq:1}),(\ref{eq:3}),(\ref{eq:5})
and (\ref{eq:6}) one can write:
\begin{eqnarray}
&&\dfrac{di_{M}}{dt}= \beta_{M}i_{M}s_{M}-\gamma_{M}i_{M}-\mu i_{M}\label{eq:9}\\
&&\dfrac{ds_{M}}{dt}= -\beta_{M}i_{M}s_{M}-\beta_{X}i_{X}\rho_{SS}-\mu s_{M} + \mu(1-v_{M}). \label{eq:10}
\end{eqnarray}
 Assuming, as above, initial herd immunity for the X-disease, i.e.  $v_{X}>v_X^{\dagger}$,  one can approximate $i_{X}\approx0$ and readily find the two steady-state solutions of Eqs.(\ref{eq:9}) and (\ref{eq:10}): a disease-free one, ${(i_{M}^{*1},s_{M}^{*1})=(0,1-v_{M})}$ and
an endemic one ${(i_{M}^{*2},s_{M}^{*2})=(R_{0}^{M} ((1-v_{M})-1)\mu/\beta_{M}, 1/R_{0}^{M})}$.
A standard linear stability analysis allows us to recover the existence of a transcritical bifurcation at $v_{M}^{\dagger}=1-1/R_{0}^{M}$ where these two fixed points exchange their stability; thus the system shifts in a continuous or smooth way from a non-propagating to an endemic state. Observe that this transition point is a natural extension of the threshold for $\mu=0$, even when no endemic state existed in that case, but just a separation between propagating and quiescent  phases.
\begin{figure}
\begin{centering}
\includegraphics{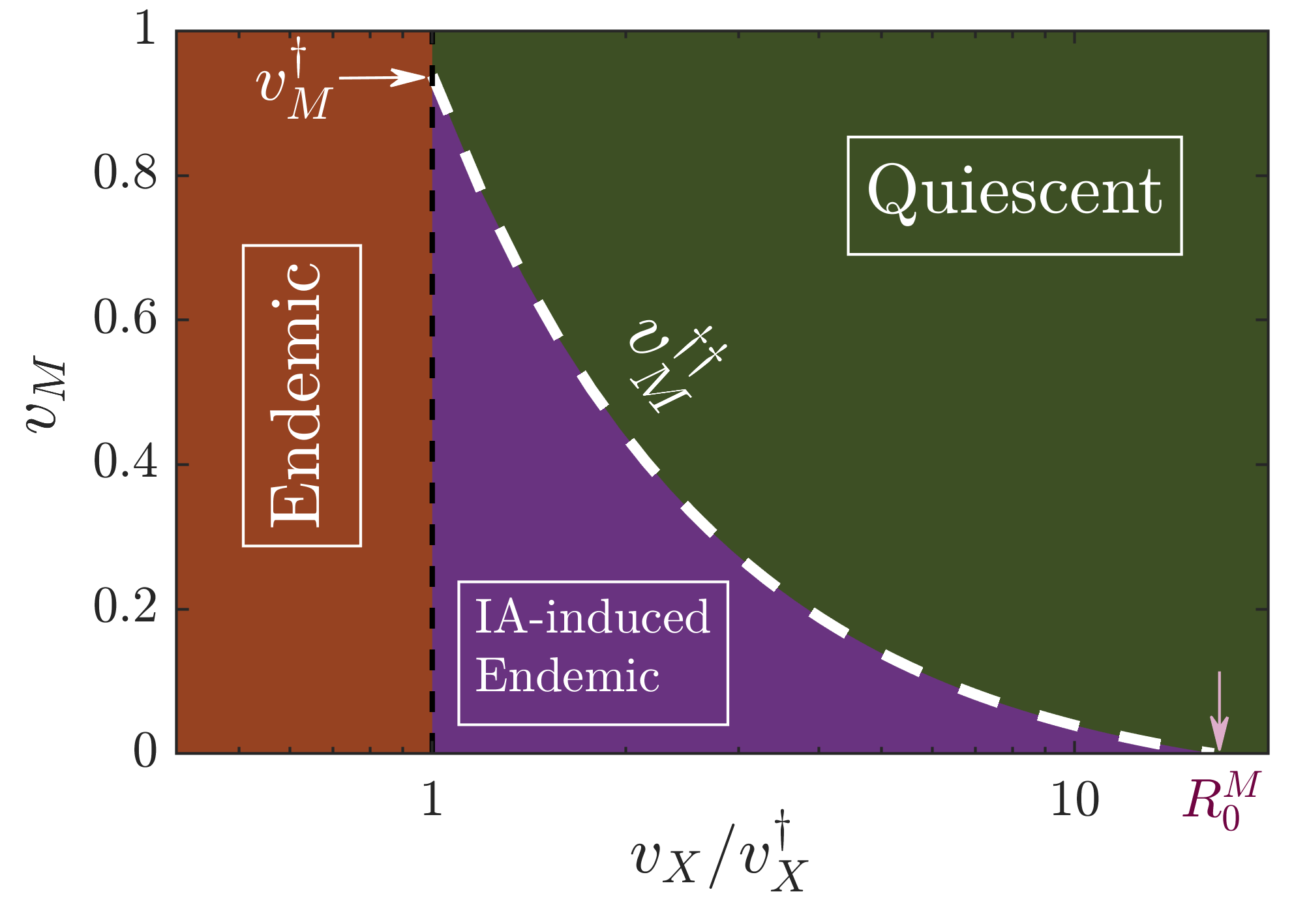}
\par\end{centering}
\caption{\textbf{Analytically determined phases for disease X in the SIR-IA model with demography}. The white-dashed line depicting the vaccination threshold under IA effects was determined through Eq.(\ref{crit_X}).
The IA-induced endemic phase is bounded by $v_{X}=v_{X}^{\dagger}$, below which the disease is always endemic, independently of IA effects (black-dashed line); and $v_{X}=v_{X}^{\dagger}R_{0}^{M}$ (marked by the pink arrow), above which the epidemics does not propagate despite the IA contribution. Notice that a triple point appears for $v_X=v_X^\dagger$ and $v_M=v_M^\dagger$ (marked by the white arrow), at which the three phases collide. 
\label{fig:Phase_Space2}}
\end{figure}
Similarly, for  the  X-disease, using the joint variables $i_X$ and $s_X$:
\begin{eqnarray}
&\dfrac{ds_{X}}{dt}=&-\beta_{X}s_{X}i_{X}-\beta_{M}\rho_{SS}i_{M}+\gamma_{M}i_{M}-\mu s_{X}\label{eq:11}\\
&&+\mu\left(1-v_{X}\right),\nonumber  \\
&\dfrac{di_{X}}{dt}=&\beta_{X}s_{X}i_{X}-\gamma_{X}i_{X}-\mu i_{X}. \label{eq:12}
\end{eqnarray}

Notice that it is not possible to solve exactly the above equations for their fixed points, as they do not form a closed set:  Eq.(\ref{eq:11}) depends on the fraction $\rho_{SS}$ of  individuals susceptible to both, measles and X-disease, as well as on the fraction $i_{M}$ of measles infectious population. However,  assuming that a steady state for measles has been reached, it is possible to show that ${\rho_{SS}^{*}=(1-v_{X})/R_0^{M}}$; using this and searching for steady-state solutions of  Eqs.(\ref{eq:11}) and (\ref{eq:12}),  one can compute the minimum value of $v_{M}$ preventing the existence of a stable X-endemic state, thus obtaining an expression for $v_{M}^{\dagger\dagger}$ (see Appendix A in the SI for a detailed derivation).

This can be further simplified  in the typical case under consideration where the immigration rate is much smaller than the measles recovery rate, ${\mu\ll\gamma_{M}}$, resulting in:
\begin{equation}
v_{M}^{\dagger\dagger}\approx\dfrac{1}{v_{X}}\left(1-\dfrac{1}{R_{0}^{X}}\right)-\dfrac{1}{R_{0}^{M}} ~= ~\dfrac{v_{X}^{\dagger}}{v_{X}}-\dfrac{1}{R_{0}^{M}}\label{crit_X}
\end{equation}
which depends not only on the  $R_{0}$'s of both diseases, but also on the vaccination coverage for disease X, as computationally observed above. Note also that the rightmost  expression  in Eq.(\ref{crit_X}) ---written in terms of $v_{X}^{\dagger}$ rather that $R_0^X$--- underlines the relationship between the two vaccination thresholds. This result is illustrated in Fig.\ref{fig:Phase_Space2}, which shows
the three resulting phases for a generic X disease: (i) disease-free, (ii) IA-induced endemic, and (iii) endemic, as well as their  phase boundaries in the ${(v_{M},v_{X}/v_{X}^{\dagger})}$ plane. 

Observe that, since we assumed ${v_{X}\geq v_{X}^{\dagger}}$, then trivially ${v_{M}^{\dagger\dagger}\leq1-\frac{1}{R_{0}^{M}}=v_{M}^{\dagger}}$ must hold,
with both critical points coinciding for ${v_{X}=v_{X}^{\dagger}}$. Similarly, as $0\leq v_{M}^{\dagger\dagger}\leq1$, one readily sees that the existence of a non-trivial second threshold $v_{M}^{\dagger\dagger}$ is limited by the constraint $v_{X}/v_{X}^{\dagger}\leq R_{0}^{M}$
(these two limits are marked with arrows in Fig.\ref{fig:Phase_Space2}). Therefore,  \emph{an immune-amnesia-induced endemic phase for X disease exists under broad conditions for realistic parameter values}.

All the above analytical results, derived from  linear stability analyses with some additional approximations, have been confirmed by numerically determining the fixed points for the full system of mean-field Eqs.(\ref{eq:1}-\ref{eq:8}), without invoking any approximation beyond numerical accuracy (see Materials and Methods). Moreover, one can also cross-check the consistency between these analytical approaches and the previous results from computational simulations, for example, by looking at Figs.\ref{fig:Bifurcation-Diagram} and \ref{fig:HeatMap}, which reveal that the analytical predictions for $v_{M}^{\dagger\dagger}$, as given by Eq.(\ref{crit_X}), explain well both the onset of the X outbreak in the deterministic calculation (Fig.\ref{fig:Bifurcation-Diagram}) and its dependence on $R_0^X$ and $v_X$ (Fig.\ref{fig:HeatMap}).

\subsection{SIR-IA model on structured networks}
After two decades of frantic activity on the development of the theory of complex networks, by now it is broadly recognized  that the structure of the underlying network of contacts plays a crucial role on spreading phenomena such as epidemics \cite{EpidemicsRandom_Mollison,Barabasi:scfree,newman_spread_2002,Pastor-Satorras,SHIRLEY2005287,keeling_networks_2005,barrat_dynamical_2008,pastor-satorras_epidemic_2001}. Thus, to have a broader view on IA effects, here we scrutinize the behavior of the SIR-IA model beyond the homogeneous-mixing approach, considering more structured topologies such as  Erd\H{o}s-Rényi (ER) random networks and power-law degree-distributed networks \cite{Erdos:1959:pmd,Barabasi:scfree,newman_structure_2003,newman_networks_2018} (see Methods).
 
 In order to make further progress, we make some simplifying assumptions: (i) Vaccination for both measles and X is considered to be performed in a random random way across the network (i.e., there is no "targeted-immunization" program selecting preferentially specific nodes for vaccination according to, e.g., their network centrality or connectivity \cite{kitsak_identification_2010,madar_immunization_2004,cohen_efficient_2003,pastor-satorras_immunization_2002}). (ii) As in the previous analyses, we impose that $v_x > v_{x}^\dagger$ to analyze how herd immunity can be potentially lost by IA effects. (iii) For simplicity, we limit ourselves here to the analytically-more-tractable non-demographic version of the SIR-IA model (i.e., $\mu=0$).  

\begin{figure}
\begin{centering}
\includegraphics{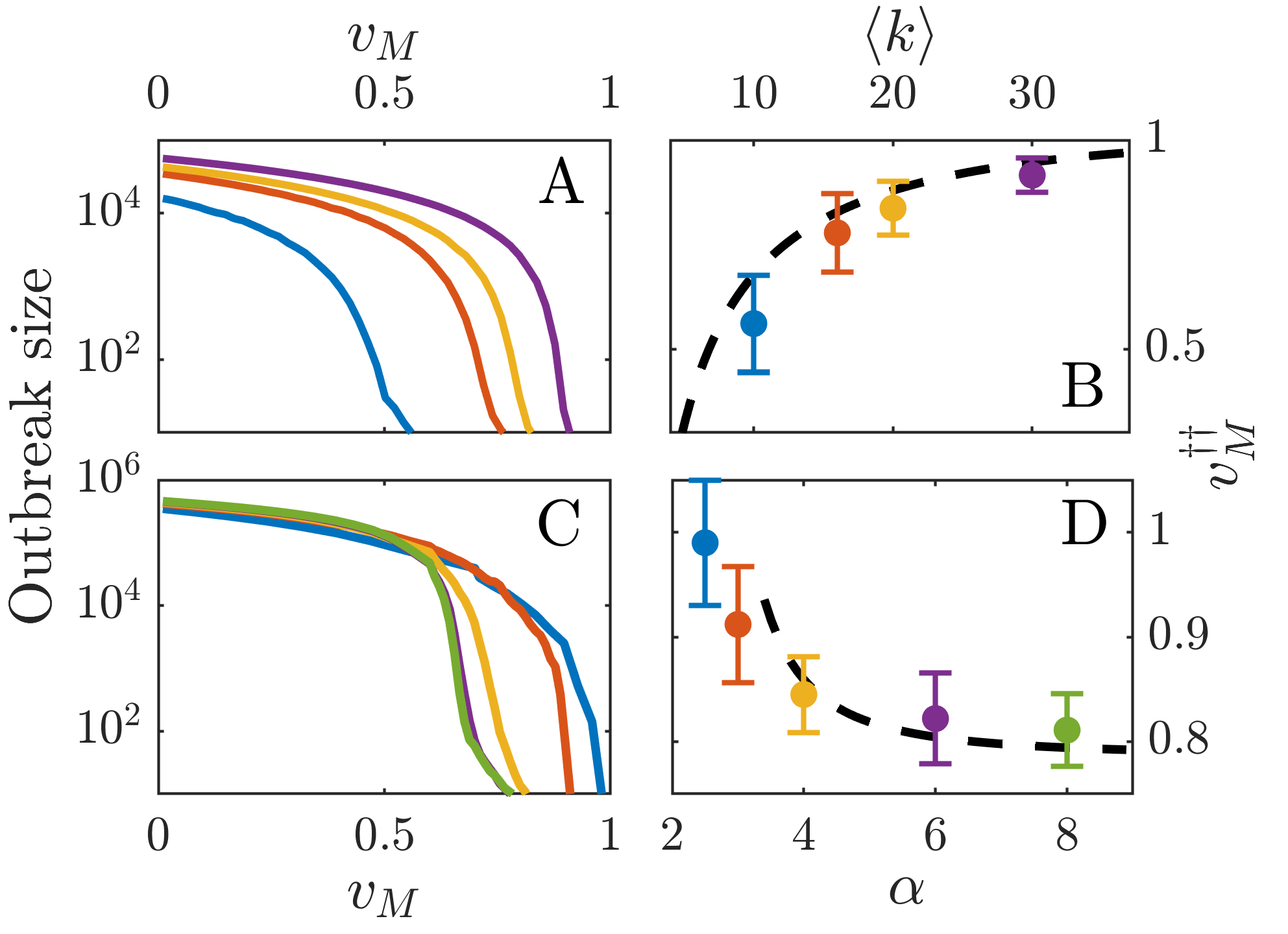}
\par\end{centering}
\caption{\textbf{Analysis of epidemic size and thresholds in Erd\H{o}s-Rényi and power-law degree distributed networks.} Outbreak sizes (as measured by the difference between X-resistant individuals before and after the introduction of one single X-infectious individual in the system) are presented for ER networks with different average degrees (\textbf{A}) and power-law networks with different degree exponents $\alpha$ (\textbf{C}). The dependence of the vaccination threshold, $v_M^{\dagger \dagger}$, with the average network degree for ER networks (\textbf{B}) and $\alpha$ exponent for  power-law networks (\textbf{D}), respectively, is compared with the corresponding analytical predictions given by the  heterogeneous-mean-field approximation. Stochastic simulations were performed in networks of size $N=10^{5}$ and $N=5\cdot10^{5}$ for ER and power-law networks, respectively. Error bars are computed as the standard deviation over $200$ realizations. The rest of parameters were chosen according to Table \ref{table1}. \label{fig:Networks}}
\end{figure}

We first report on computational findings for stochastic (Gillespie) simulations of the SIR-IA model on structured networks. Let us remark that, in order to compare ER networks with different average connectivity (or "degree") $\langle k \rangle$, we defined an \textit{infectivity per contact} $\beta_0$ so that $\beta=\beta_0\langle k \rangle$, with $\beta_0^X=0.017$ and $\beta_0^M=0.14$. Within this convention, the values presented in Table \ref{table1} are recovered for an average network degree $\langle k \rangle \approx 20$. These same values of  $\beta_0^X$ and $\beta_0^M$ where also used in power-law degree distributed networks, but in this case the average connectivity ${\left\langle k\right\rangle =u ({1-\alpha})/({2-\alpha})}$ was kept fixed by changing the minimum node degree $u$ according to the chosen exponent $\alpha$. 

Fig.\ref{fig:Networks}A shows results of the mean epidemic size
for X-disease outbreaks occurring on ER networks with different average connectivity, as a function of the measles vaccination coverage $v_M$. It can be readily noticed that the vaccination threshold $v_M^{\dagger \dagger}$ grows and the transition becomes sharper as $\langle k \rangle $ is increased. Remarkably, the transition becomes very abrupt for large mean degrees, implying that a small variation in the measles vaccination level can induce a dramatic effect  on the typical size of the subsequent X-disease outbreaks.  For instance, for $\langle k \rangle \geq 30 $, lowering the vaccination coverage from $v_M=0.9$ to $v_M=0.82$ (a reduction on the number of vaccinated individuals of just an 8\% of the population size) entails an increase of two orders of magnitude in the subsequent X-outbreaks.

On the other hand, for power-law degree distributions $p_{k}\sim k^{-\alpha}$,  numerical simulations reveal that reducing $\alpha$ leads to larger values of $v_M^{\dagger \dagger}$, as illustrated in Fig.\ref{fig:Networks}C. The Figure also shows that, in this case, the transition becomes more abrupt as larger values of $\alpha$ are considered. At the light of these result, one could wonder whether the transition could eventually become discontinuous or \emph{"explosive"} for larger values of $\langle k \rangle $ or $\alpha$, as it has been shown to occur in other models with cooperative contagion \cite{grassberger_phase_2016,cui_mutually_2017}. 

Remarkably, the dependence of the vaccination threshold on the average network degree and exponent $\alpha$ can be described using the \textit{heterogeneous mean-field}  approach, a generalization  of the mean-field theory that groups all nodes with the same  degree into a common class \cite{pastor-satorras_epidemic_2001}, which has been shown to be accurate (up to finite size corrections) when applied to epidemic models that do not posses non-trivial steady states \cite{castellano_thresholds_2010}. For the SIR-IA model, the calculations turn out to be a mathematical "tour the force", so the details are deferred to Appendix B in the SI, and here we just present  the final results for both ER and power-law degree-distributed networks.

In particular, for ER networks, calling  $q=s_1^M(T)/(1-v_M)$ to the fraction of degree-one M-susceptible individuals at the time $T$ of inserting an X-infectious seed with respect to the initially M-susceptible population fraction, we found the following equation: 
\begin{equation}
q (1-v_{M}) e^{\left\langle k\right\rangle \left(q-1\right)}-\log\left(q\right)/R_{M}^{eff}-\left(1-v_{M}\right)=0,\label{relation_q&vm}
\end{equation}
where $R_{M}^{eff}=\beta_0^{M}/\gamma_M$. Defining the auxiliary function
${\zeta(q)=q v_{X}\left(1+\left\langle k\right\rangle q\right)e^{-\left\langle k\right\rangle \left(1-q\right)}}$
we were able to derive a closed equation for the vaccination threshold
as a function of parameter values:
\begin{equation}
v_{M}^{\dagger\dagger}(q)=\frac{1/R_{X}^{eff}-\left(1+\left\langle k\right\rangle \right)+\zeta(q)}{\zeta(q)-v_{X}\left(1+\left\langle k\right\rangle \right)}. 
\end{equation}
This expression, when inserted into Eq.(\ref{relation_q&vm}), can be numerically solved, leading to an analytical determination of $v_{M}^{\dagger\dagger}$. Fig.\ref{fig:Networks}B shows such a theoretical solution (black-dashed line) as a function  of the network average degree, together with the computational results obtained by averaging over many runs of the stochastic simulations of the model running on ER networks with different  averaged connectivities.

On the other hand, for power-law degree-distributed networks with $\alpha>3$ we defined $\xi=log\left(\frac{1-v_{M}}{s_{u}^{M}}\right)$ ---where $u$ is now the minimum degree of a node in the network--- and ${v_{X}^{\dagger}=1-\frac{\left\langle k\right\rangle}{R_{X}^{eff}\left\langle k^{2}\right\rangle}}$, which is the corresponding threshold in a standard SIR model with vaccination \cite{pastor-satorras_immunization_2002}. In terms of these quantities, we obtained (see Appendix B.3),
\begin{equation}
v_{M}^{\dagger\dagger}=\dfrac{\left(\alpha-3\right)\xi^{\alpha-3}\Gamma(3-\alpha,\xi)-v_{X}^{\dagger}/v_{X}}{\left(\left(\alpha-3\right)\xi^{\alpha-3}\Gamma(3-\alpha,\xi)-1\right)}.
\end{equation}
with the constraint
\begin{equation}
\left(1-v_{M}\right)\left(e^{-\xi}-(\alpha-2)\xi^{\alpha-2}\Gamma\left(2-\alpha,\xi\right)-1\right)+\dfrac{\gamma_{M}}{u\beta_{M}^{0}}\xi=0
\end{equation}
needed to determine $\xi$.
Conversely, for scale-free networks (i.e., $2<\alpha \leq 3$) one finds that outbreaks do not propagate only if  $v_{M}=v_{X}=1$, i.e., the whole population needs to be vaccinated of both diseases to prevent the epidemic. This result is in consonance with the well-known phenomenon of vanishing epidemic threshold in BA networks \cite{barabasi2016network,cui_mutually_2017, may_infection_2001, peng_vaccination_2013}: all single nodes need to be vaccinated to prevent epidemics propagation, a result that stems from the existence of super-spreaders, i.e. network hubs. However, our analytical results predict also that, for the SIR-IA model, the M-vaccination threshold for the propagation of an X-disease outbreak depends also on the X-vaccination coverage and, hence, \emph{the vaccination threshold can saturate even in scale-rich networks, with $\alpha>3$} (see Fig.\ref{fig:Networks}D, where for $v_X=0.9$ a minimum M-vaccination level preventing an X outbreak is defined only in networks with $\alpha\gtrsim3.4$). 
Nevertheless, it  is important to remark  that these result are strictly valid only  in the  infinite-size limit ($N\rightarrow\infty$), in which the second-moment of the degree distribution truly diverges in scale-free networks. In fact, for $\langle k \rangle = 15$ and $\alpha=3$, we found  a vaccination threshold clearly below $1$ in stochastic simulations (see Fig.\ref{fig:Networks}D). This phenomenon that should come as no surprise, since finite size effects in the SIR model have been shown to be responsible for the appearance of non-trivial thresholds  in scale-free networks at sufficiently low transmission rates \cite{may_infection_2001}. 

As shown in Figs.\ref{fig:Networks}B and \ref{fig:Networks}D, the analytical results  match quite well the values of $v_M^{\dagger\dagger}$ found in stochastic simulations.  In both ER and power-law networks, small disagreements with computational results are most likely rooted in \emph{finite-size effects} and the associated possibility of stochastic fade-out that occurs in finite networks and not in infinitely large ones; however, performing a proper finite-size analysis to study such effects is beyond the scope of the present work \cite{van_herwaarden_stochastic_1997,meerson_wkb_2009,hartfield_introducing_2013}.

Finally, to the question of whether discontinuous transitions could be found in power-law networks with sufficiently high values of $\alpha$, application of the heterogeneous-mean-field theory to our model predicts that ---provided a non-trivial vaccination threshold exist in the large-$N$ limit--- transitions between a quiescent and a propagating phase are always continuous independently of the $\alpha$ exponent (see Appendix B.3) and a similar conclusion holds for ER networks even in the limit of large average connectivities. Still, it remains to be carefully investigated how other structural aspects such as clustering, geography, and network modularity, to name but a few, could affect such a conclusion.

\section{Conclusions}
We have seen that, when measles vaccination policies are relaxed, the expected  herd-immunity for any secondary infectious disease X can be lost owing to the proliferation of individuals affected by immune-amnesia. In particular, under IA effects, the epidemic threshold is shifted so that severe outbreaks can take place even under extensive X vaccination.  We have studied the conditions under which measles vaccination can prevent such outbreaks. To support the generality of our findings, we considered two different variants of the SIR-IA model: one in which all the state transitions are given by infectious/recovery processes, and a second version in which demographic effects are also taken into account. For both models, we were able to derive ---under homogeneous-mixing assumptions--- analytical expressions for the epidemic threshold in terms of the fraction of people vaccinated of measles, which were able to reproduce with significant accuracy the results obtained through simulations. Remarkably, the analytical results also allowed us to construct a full phase-diagram in both models, where three distinct phases were found: quiescent, endemic/propagating and, more remarkably, IA-induced endemic/propagating phases.

We have also studied the persistence of the X disease under IA when more realistic network architectures ---beyond the homogeneously-mixed population paradigm, were considered. In particular, when the SIR-IA model is implemented in random ER networks, it was shown that larger fractions of M-vaccination coverage are necessary to prevent outbreaks as the average network connectivity increases. We remark that this dependence is highly non-linear (as also shown by our analytical results), with the epidemic threshold decaying very fast at low connectivity values. Thus, when fighting an outbreak in ER networks under IA effects, it is likely that measures taken to lower the average connectivity, even without imposing full confinement ---e.g., limiting the allowed number of individuals in gatherings--- can have indeed a profound impact in the spreading of the disease.

On the other end of the spectrum, for scale-free networks the vaccination thresholds is equal to unity, implying
that the whole population needs to be vaccinated of both diseases to prevent epidemics propagation; this is the counterpart of the well-known phenomenon of vanishing epidemic threshold, and is predicted by our analytical calculations in the limit of very large networks. The fact that outbreaks persist even when a large percentage of the population is vaccinated manifests not only the key role of hubs (i.e. super-spreaders) in the spreading process, but also the scarce effectiveness of vaccine uptake measures when these are randomly administrated (as opposed to, e.g., targeting  the most connected nodes \cite{pastor-satorras_immunization_2002,madar_immunization_2004}). Aiming for a more profound understanding of this effect, we considered general power-law degree-distributed networks which drift from the scale-free to the random regime by considering values of $\alpha>3$.  As shown by stochastic computer simulations, the vaccination threshold becomes non-trivial  once the network presents a non-divergent second moment, scaling with the exponent $\alpha$ as predicted by the analytically derived expression in the limit of $N\rightarrow\infty$. 

Let us  finally discuss the nature of the transition between the propagating and quiescent regimes. It was shown in \cite{chen_outbreaks_2013} and further investigated in \cite{cai_avalanche_2015,janssen_first-order_2016,grassberger_phase_2016,cui_mutually_2017} , that cooperative epidemics can show hybrid-type phase transitions for large enough $\alpha$ in the limit of large system sizes. Although performing a detailed analysis of this important issue is beyond our scope here (and is left for a further work), the implications of a possible discontinuous transition are enormous from the dynamical perspective, opening the door to catastrophic regime shifts \cite{catastrophic1,catastrophic2}: under such conditions ---with just a slight reduction in the fraction of vaccinated population--- the system could suddenly undergo a 
transition from a quiescent state with overall herd immunity to a state in which anomalously large pandemics could surge 
As discussed for the homogeneous-mixing case, at the epidemic threshold we find a second-order phase transition, which holds at first sight when more realistic networks are considered. Nevertheless, we see that the transition becomes more abrupt as we consider larger mean degrees in ER networks, or greater exponents $\alpha$ in power-law degree distributed networks. For the latter, however, application of the heterogeneous-mean-field theory to our model predicts a continuous phase transition independently of the value of $\alpha$. More realistic network architectures including clustering, modularity or embedding into a metric space, as well as possibly temporally changing networks, still need to be analyzed to have a full
view of the potential effects of immune amnesia on real populations. 
Furthermore, the possibility of finding actual first-order transitions when further cooperative interactions are implemented is left for future work. In particular, let us notice  that during lock-down periods to fight COVID-19, vaccination coverage for measles decreases, thus creating a feedback loop that could enormously enhance the impact of a future M-outbreaks and thus of immune amnesia.

Our results not only pave the way to the study of cooperative contagions from the perspective of immunization (in comparison with other works, in which cooperativity is modeled as changes in infectivity or recovery rates \cite{chen_outbreaks_2013,sanz_dynamics_2014,janssen_first-order_2016}), but also open a branch of new exciting questions. For example, it is known that vaccination strategies targeting the hubs in scale-free networks can effectively reduce to finite values the epidemic threshold \cite{dezso_halting_2002}. It would be therefore very relevant to study the effects that using different immunization strategies for each disease (e.g., random and targeted vaccination) have on the existence or absence of an epidemic threshold in scale-free networks. 

Let us also emphasize that we have developed an advanced
version of the heterogeneous-mean-field approach which
can be applied to derive analytical results in similar
problems of cooperative contagion or, more in general, 
when different epidemics coexist, influencing each other.

It is our hope that this work makes it clear the importance of keeping measles vaccination (and vaccination in general) at levels at high as possible, to prevent immune-amnesia effects to have a strong negative impact at a global level. We also hope that this work fosters further
investigations along these lines as well as novel developments in other directions taking advantage of  the  techniques we have set up.

\section*{Matherials and Methods}
The steady-state solutions reported in Figs.\ref{fig:Minimum-fraction-of} and \ref{fig:HeatMap} were obtained by solving Eqs.(\ref{eq:1}-\ref{eq:8}) for their fixed points. The eigenvalues of the associated Jacobian matrix were then analyzed to determine their possible stability \cite{strogatz_nonlinear_2000}. Only the resulting  stable fixed points are plotted in such figures. Likewise, the deterministic trajectories in the standard mean-field approximation shown in Fig.\ref{fig:Minimum-fraction-of} (insets) were obtained by integrating Eqs.(\ref{eq:1}-\ref{eq:8}) with Matlab ode23 function, which implements an explicit Runge-Kutta (2,3) pair algorithm \cite{Bogacki_Rhunge}. On the other hand, simulations of the stochastic dynamics were performed through the standard Gillespie algorithm in the homogeneous
mean-field approach \cite{Gillespie} and through a network-adapted Gillespie for any other network structure, as described in \cite{kiss_mathematics_2017}. Unless otherwise specified, initial conditions for the simulations were set in agreement with the vaccination rates following $\rho_0=\{\rho_{SS}=(1-v_{M})(1-v_{X})-\left(\varepsilon_{X}+\varepsilon_{M}\right)$,
$\rho_{SI}=\varepsilon_{X}$, $\rho_{SR}=0$, $\rho_{IS}=\varepsilon_{M}$, $\rho_{IR}=0$, $\rho_{SR}=(1-v_{M})v_{X}$, $\rho_{RS}=v_{M}(1-v_{X})$, $\rho_{RR}=v_{M}v_{X}\}$. To obtain the deterministic solutions (Figs.\ref{fig:Minimum-fraction-of} and \ref{fig:HeatMap}) we set   $\varepsilon_{M}=\varepsilon_{X}=50/N$, while for the the stochastic simulations we chose $\varepsilon_{M}=1/N$ and $\varepsilon_{X}=0$, introducing one X-infectious individual only after the measles outbreak has fade out. In all cases the total population size was fixed to $N=10^{5}$.

In what respect structured networks, Erd\H{o}s-Rényi graphs were constructed following the standard algorithm as described in \cite{Erdos:1959:pmd}, while for power-law degree-distributed networks we used the configuration model \cite{bornholdt_handbook_2006} to generate a first graph from which we then removed all multiple and self-connections. Although this last step may introduce correlations within the networks, they are negligible for all purposes when $\alpha>3$ and large system sizes are considered \cite{catanzaro_generation_2005}.

\section*{Acknowledgements}
MAM acknowledges the Spanish Ministry and Agencia Estatal de investigaci{\'o}n (AEI) through grant FIS2017-84256-P (European Regional Development Fund), as well as the Consejería de Conocimiento, Investigación  Universidad, Junta de Andalucía and European Regional Development Fund, Ref. A-FQM-175-UGR18 and SOMM17/6105/UGR for financial support.  We thank V. Buendia and R. Burioni for valuable discussions.

\section*{Supplementary Information}

\subsection*{Appendix A: SIR-IA model with demography in homogeneous networks}

We begin the analytical study by assuming that, initially, the number of X-infectious individuals is very low due to a high-vaccination coverage $v_{X}\lesssim1$.  Under this approximation, Eqs.(2) and (3) of the main text describing the evolution of the total fraction of susceptibles and infectious of measles, can be written as:
\begin{equation}
\begin{aligned}
\dfrac{di_{M}}{dt}= & \beta_{M}i_{M}s_{M}-\gamma_{M}i_{M}-\mu i_{M},\\
\dfrac{ds_{M}}{dt}= & -\beta_{M}i_{M}s_{M}-\mu s_{M}+\mu\left(1-v_{M}\right).
\end{aligned}
\end{equation}
In this way, the dimensionality of the problem is largely reduced, allowing us to compute the stationary states in the now independent set of variables $\{s_{M},i_{M}\}$: 
\begin{equation}
\begin{aligned}
\xi_{1}=& \left(i_{M}^{*1},s_{M}^{*1}\right)=\left(0,1-v_{M}\right),\\
\xi_{2}=& \left(i_{M}^{*2},s_{M}^{*2}\right)=\left(\dfrac{\mu\left(R_{0}^{M}\left(1-v_{M}\right)-1\right)}{\beta_{M}},\dfrac{1}{R_{0}^{M}}\right).
\end{aligned}
\end{equation}
To analyze the stability of the above fixed points, we compute the Jacobian associated with the linearization of Eqs.(2) and (3) around a point in which there is no infectious individuals of X (i.e. $i_{X}=0$), as we assumed to be above the herd-immunity threshold for this disease. Therefore, linearizing around any point $\left(i_{M}^{*},s_{M}^{*},i_{X}=0\right)$ we have:
\begin{equation}
J_{M}=\left(\begin{array}{cc}
\beta_{M}s_{M}^{*}-(\gamma_{M}+\mu) & \beta_{M}i_{M}^{*}\\
-\beta_{M}s_{M}^{*} & -\beta_{M}i_{M}^{*}-\mu
\end{array}\right).   
\end{equation}
If $v_{M}$ is chosen as the control parameter, one can perform a stability analysis by evaluating the determinant ($\Delta$) and the trace ($\tau$) of $J_{M}$ at the each of the fixed points \cite{strogatz_nonlinear_2000}. Thus, the disease-free fixed point $\xi_{1}$ is a stable focus of the linearized, X-disease-free system  if $v_{M}>1-1/R_{0}^{M}$ , and a saddle-node otherwise.  On the other hand, the endemic stationary state $\xi_{2}$ does only exist  provided $v_{M}<1-1/R_{0}^{M}$, and it is always a stable focus. Therefore, the epidemic threshold for the propagation of measles is  given by  the critical value  $v_{M}^{\dagger}=1-1/R_{0}^{M}$.  

Now, to study the epidemic threshold for X using a similar approach requires of some approximations. In particular, we assume that a stationary state for the measles variables $s_{M}$ and $i_{M}$ is
reached (regardless if endemic or quiescent) before the evolution
of disease X can significantly progress. This a reasonable assumption
because, on one hand, the dynamics of measles occurs at a faster time
scale (shorter recovery period and faster infection rate). Moreover, we start with an initial state for X that is over
herd-immunity, and thus activity (in terms of the fraction of X infectious individuals)
is very low until IA effects can convert a large fraction of resistant individuals into susceptibles. Within this approximation it is possible to solve Eqs.(4) and (5) in the main text for the stationary states, obtaining:

\begin{equation}
\begin{aligned}
\left(i_{X}^{*},s_{X}^{*}\right)=&\left(0,\dfrac{i_{M}^{*}}{\mu}\left(\gamma_{M}-\beta_{M}\rho_{SS}^{*}\right)+\left(1-v_{X}\right)\right),\label{General_FP}\\
\left(i_{X}^{*},s_{X}^{*}\right)=&\left(\frac{i_{M}^{*}\left(\gamma_{M}-\beta_{M}\rho_{SS}^{*}\right)+\mu\left(1-v_{X}-\frac{1}{R_{0}^{X}}\right)}{\gamma_{X}+\mu},\frac{1}{R_{0}^{X}}\right),
\end{aligned}
\end{equation}  
where $i_{M}^{*}$ and $\rho_{SS}^{*}$ denote the steady-state fraction of M-infectious and fully-susceptible individuals, respectively.

Beginning with a scenario where the system is in a disease-free stationary state $\xi_{1}^{M}$ for measles, we find:
\begin{equation}
\begin{aligned}
\psi_{1}=&\left(i_{X}^{*1},s_{X}^{*1}\right)=\left(0,1-v_{X}\right),\\
\psi_{2}=&\left(i_{X}^{*2},s_{X}^{*2}\right)=\left(\frac{\mu}{\beta_{X}}\left(R_{0}^{X}\left(1-v_{X}\right)-1\right),\frac{1}{R_{0}^{X}}\right).
\end{aligned}    
\end{equation}
Notice that these stationary states are the X-counterparts of $\xi_{1}$
and $\xi_{2}$. This symmetry reflects the fact that, without
a preceding outbreak of measles and its consequent IA effects, the
evolution of the X infection should follow that of an independent SIR model with vaccination and demographic
dynamics, where the endemic state $\psi_{2}$ is stable provided $v_{X}<v_{X}^{\dagger}=1-1/R_{0}^{X}$.

Let us move now to the more interesting case in which there is a endemic measles stationary state, given by $\xi_{2}$. Since now Eqs.(4) and (5) do not form a closed set, we first
proceed to estimate $\rho_{SS}$ using $s_{M}=\rho_{SS}+\rho_{SR}$ and Eq.(1c). Assuming stationarity in the number of M-infectious individuals and low initial fraction of X-infectious ($i_{X}\approx0$), one can write:
\begin{equation}
\dfrac{d\rho_{SR}}{dt}\approx\,-\beta_{M}\rho_{SR}i_{M}^{*2}-\mu \rho_{SR}^{*}+\mu v_{X}(1-v_{M})=0,
\end{equation}
from where $\rho_{SR}^{*}=\frac{v_{X}}{R_{0}^{M}}$ and $\rho_{SS}^{*}=\frac{1-v_{X}}{R_{0}^{M}}$. It is important to remark that these are not truly stationary states, but serve us as an estimation of the expected distribution of M-susceptibles
near the measles endemic state $\xi_{2}$ before the onset of an X outbreak. Let us nevertheless carry on with the analysis and impose fixed values $\rho_{SS}^{*}$ and $i_{M}^{*2}$ in Eq.(\ref{General_FP}). One then obtains the following pair of stationary states:
\begin{equation}
\begin{aligned}
\psi_{3} =& \left(i_{X}^{*3},s_{X}^{*3}\right)=
\left(0,\dfrac{\gamma_{M}}{\beta_{M}}\left(R_{0}^{M}\left(1-v_{M}\right)-1\right)+\left(1-v_{X}\right)\left(v_{M}+\dfrac{1}{R_{0}^{M}}\right)\right),\\
\psi_{4} =& \left(i_{X}^{*4},s_{X}^{*4}\right)= \left(\dfrac{\mu}{\mu+\gamma_{X}}\left(s_{X}^{*3}-\dfrac{1}{R_{0}^{X}}\right),\dfrac{1}{R_{0}^{X}}\right).
\end{aligned}    
\end{equation}

As before, one can linearize the above system around each of these solutions. Studying the determinant and trace of the resulting Jacobian matrix:
\begin{equation}
J_{X}=\left(\begin{array}{cc}
-\mu-\beta_{X}i_{X}^{*} & -\beta_{X}s_{X}^{*}\\
\beta_{X}i_{X}^{*} & \beta_{X}s_{X}^{*}-\left(\gamma_{X}+\mu\right)
\end{array}\right),
\end{equation}
allow us to conclude that the X-disease-free steady state is a stable focus if $s_{X}^{*3}<\frac{1}{R_{0}^{X}}$, and a saddle-node otherwise; whereas $\psi_{4}$ is a stable focus only when $s_{X}^{*3}>\frac{1}{R_{0}^{X}}$ and does not exist otherwise. Therefore, writing the explicit form found for $s_{X}^{*3}$ and taking $v_{M}$ as the control parameter, one finally obtains the epidemic threshold for the X disease in terms of the fraction of M-vaccinated population:
\begin{equation}
v_{M}^{\dagger\dagger}=\dfrac{\dfrac{1}{R_{0}^{X}}-\left(1-v_{X}\right)\dfrac{1}{R_{0}^{M}}+\dfrac{\gamma_{M}}{\beta_{M}}\left(1-R_{0}^{M}\right)}{1-v_{X}-\dfrac{R_{0}^{M}\gamma_{M}}{\beta_{M}}}\label{crit_X_Full}
\end{equation}
In the typical case in which $\mu \ll \gamma_M$, one can approximate $\gamma_M/\beta_M \approx 1/R_{0}^{M}$, and the vaccination threshold presented in Eq.(10) of the main text is recovered:
\begin{equation}
v_{M}^{\dagger\dagger}\approx\dfrac{1}{v_{X}}\left(1-\dfrac{1}{R_{0}^{X}}\right)-\dfrac{1}{R_{0}^{M}} ~= ~\dfrac{v_{X}^{\dagger}}{v_{X}}-\dfrac{1}{R_{0}^{M}}\label{crit_X_SI}
\end{equation}

\subsection*{Appendix B:  SIR-IA model without demography in heterogeneous networks.}

The Heterogeneous mean-field (HMF) approach \cite{pastor-satorras_epidemic_2001} is a degree-block approximation carried out by assuming that nodes with the same degree $k$ are statistically
equivalent, i.e., belong to the same connectivity class. Within this framework, the total fraction of X-infectious individuals at each degree-class, $i_{k}^{X}$, obeys:
\begin{equation}
\dfrac{di_{k}^{X}(t)}{dt}=\beta_{0}^{X}ks_{k}^{X}\theta_{X}-\gamma_{X}i_{k}^{X},
\end{equation}
where $\theta_{X}=\dfrac{1}{\left\langle k\right\rangle }\sum_{k}kp_{k}i_{k}^{X}$
is the density function, defined as the fraction of infected nodes
in the neighborhood of a susceptible node with degree $k$, and can
be shown to be independent of $k$ if the network has no degree-correlations
\cite{barabasi2016network}. Multiplying both sides of the above equation by the
excess degree, $kp_{k}/ \left\langle k\right\rangle$, and
summing over all possible values of $k$, we get an equation for the
evolution of the density function for X-infectious:
\begin{equation}
\dfrac{d\theta_{X}}{dt}=\frac{1}{\left\langle k\right\rangle }\beta_{0}^{X}\left\langle k^{2}s_{X}\right\rangle \theta_{X}-\gamma_{X}\theta_{X}.
\end{equation}
Placing a single X-infectious individual at time $T$, the
condition for an outbreak to spread in the mean-field approximation
is trivially given by  $\dfrac{d\theta_{X}(T)}{dt}>0$. Hence, the epidemic threshold is determined by the condition: 
\begin{equation}
\left\langle k^{2}s_{X}(T)\right\rangle =\dfrac{\left\langle k\right\rangle }{R_{X}^{eff}}.\label{ep_th1}
\end{equation}
where $R_{X}^{eff}=\beta_{0}^{X}/\gamma_{X}$ is an effective
$R_{0}$ per contact. To estimate the critical
value $\left\langle k^{2}s_{X}(T)\right\rangle $ for which the epidemic
breaks out, we need an expression for the minimum fraction of X-susceptible individuals
in each degree-class at the time $T$ of inserting the X-seed:
\begin{equation}
s_{k}^{X}(T)=\rho_{k}^{SS}(T)+\rho_{k}^{RS}(T).
\end{equation}
Using the same approximation as in the case with $\mu\neq0$, we
consider that an outbreak of measles propagates and dies out before
a seed of X-infectious individuals is placed in the system at time
$T$ (i.e. $i_{k}^{M}(t)=0, \forall t \geq T$ and $i_{k}^{X}(t)=0, \forall t<T$). If this is the case, then $\rho_{RS}$ is a "dead-end" state for $t\leq T$, and three sources contribute to its value at $t=T$:
\begin{itemize}
\item Initial fraction of M-resistants, that are not resistant for X: $\rho_{k}^{RS}(0)=v_{M}(1-v_{X})$
\item Fully susceptible individuals that became resistant for measles after
undergoing an infection ($\rho_{SS}\rightarrow\rho_{RS}$): $\rho_{k}^{SS}(0)-\rho_{k}^{SS}(T)=(1-v_{M})(1-v_{X})-\rho_{k}^{SS}(T)$
\item X-resistants who became infected with measles, and lost their immunity
due to IA effects ($\rho_{SR}\rightarrow\rho_{RS}$): $\rho_{k}^{SR}(0)-\rho_{k}^{SR}(T)=(1-v_{M})v_{X}-\rho_{k}^{SR}(T)$
\end{itemize}
Hence, summing up all contributions one obtains:
\begin{equation}
s_{k}^{X}(T)=1-v_{M}v_{X}-\rho_{k}^{SR}(T).\label{sX_vs_rhoSR}
\end{equation}
The epidemic threshold for X depends therefore on the fraction $\rho_{k}^{SR}(T)$
of M-susceptible individuals who are X-resistant at the end of the
measles outbreak. During the latter, $i_{k}^{X}=0$ $\forall k$,
and thus we can write:
\begin{equation}
\begin{aligned}
\dfrac{d\rho_{k}^{SS}(t)}{dt}=-\beta_{0}^{M}k\rho_{k}^{SS}\theta_{M},\\
\dfrac{d\rho_{k}^{SR}(t)}{dt}=-\beta_{0}^{M}k\rho_{k}^{SR}\theta_{M}.
\end{aligned}
\end{equation}
Dividing the above equations and integrating:
\begin{equation}
\rho_{k}^{SR}\left(T\right)=\left(\dfrac{\rho_{k}^{SR}(0)}{\rho_{k}^{SS}(0)}\right)\rho_{k}^{SS}(T).
\end{equation}
Now, since vaccination is randomly performed with the same probability across all nodes, we have $\rho_{k}^{SR}(0)=\left(1-v_{M}\right)v_{X}$
and $\rho_{k}^{SS}(0)=\left(1-v_{M}\right)\left(1-v_{X}\right)$ $\forall k$,
whence:
\begin{equation}
\begin{aligned}
\rho_{k}^{SR}(T) =\dfrac{v_{X}}{1-v_{X}}\rho_{k}^{SS}(T)=\dfrac{v_{X}}{1-v_{X}}\left(s_{k}^{M}(T)-\rho_{k}^{SR}(T)\right)=v_{X}s_{k}^{M}(T).
\end{aligned}
\end{equation}
This relation between $\rho_{k}^{SR}(T)$ and $s_{k}^{M}(T)$ allows
us to finally write Eq.(\ref{sX_vs_rhoSR}) as an expression linking
the fraction of X and M-susceptibles at time $T$:
\begin{equation}
s_{k}^{X}(T)=1-v_{M}v_{X}-v_{X}s_{k}^{M}(T).\label{XandMsus}
\end{equation}
It follows then that the value $v_{M}^{\dagger\dagger}$ at which the
epidemic threshold takes place, according to Eq.(\ref{ep_th1}), is given by:
\begin{equation}
\left\langle k^{2}\right\rangle \left(1-v_{M}^{\dagger\dagger}v_{X}\right)-v_{X}\left\langle k^{2}s_{M}(T)\right\rangle =\dfrac{\left\langle k\right\rangle }{R_{X}^{eff}}.\label{Crit_General}
\end{equation}
All it remains now is to determine the fraction of M-susceptibles
at the end of the measles outbreak. To do so, we follow a similar approach to the one presented in \cite{lucas_exact_2012} for a standard SIR model. Let us define the minimum
degree of any node in a network by $u=min(k)$ and write the equation
for the rate of change of M-susceptible individuals in the class of degree $k$:
\begin{equation}
\dfrac{ds_{k}^{M}}{dt}=-\beta_{0}^{M}ks_{k}^{M}\theta_{M}.
\end{equation}
Noticing that $\dfrac{ds_{k}^{M}}{ds_{u}^{M}}=\dfrac{ks_{k}^{M}}{us_{u}^{M}}$, one can then write:
\begin{equation}
s_{k}^{M}(t)=C_{0}\left(s_{u}^{M}(t)\right){}^{\frac{k}{u}}.
\end{equation}
with $C_{0}(k)=\dfrac{s_{k}^{M}(0)}{\left(s_{u}^{M}(0)\right){}^{k/u}}=\left(1-v_{M}\right)^{1-\frac{k}{u}}$.
Therefore, the fraction of susceptible individuals for each degree class at the end of the measles outbreak can be expressed as:
\begin{equation}
s_{k}^{M}(T)=\left(1-v_{M}\right)\left(\frac{s_{u}^{M}(T)}{1-v_{M}}\right)^{k}.\label{sk_vs_sm}
\end{equation}
Notice how, with the above relation, it is possible to determine the
fraction of M-susceptible individuals at any time and degree-class $k$ just
by studying the minimum-degree class. We introduce now the variable
$z(t)=-\log\left(s_{u}^{M}(t)\right)$, so that
\begin{equation}
\dot{z}=-\dfrac{\dot{s}_{u}^{M}(t)}{s_{u}^{M}(t)}=\beta_{M}^{0}u\theta_{M}.\label{eq:z_dot1}
\end{equation}
In terms of this new quantity and using Eq.(\ref{sk_vs_sm}), one can
write the density function for M-infectious nodes as:
\begin{equation}
\begin{aligned}
\dot{\theta}_{M} =\dfrac{1}{\left\langle k\right\rangle }\sum_{k}kp_{k}\frac{di_{k}^{M}}{dt}=\dfrac{\beta_{M}^{0}}{\left\langle k\right\rangle }\sum_{k}p_{k}k\text{\texttwosuperior}C_{0}(k)\left(s_{u}^{M}\right)^{\frac{k}{u}}\theta_{M}-\gamma_{M}\theta_{M}=\theta_{M}\left(\dfrac{\beta_{M}^{0}}{\left\langle k\right\rangle }\sum_{k}p_{k}k^{2}C_{0}(k)e{}^{-\frac{kz}{u}}-\gamma_{M}\right).
\end{aligned}
\end{equation}
 Dividing by Eq.(\ref{eq:z_dot1}):
\begin{equation}
\dfrac{d\theta}{dz}=\dfrac{\sum_{k}p_{k}k\text{\texttwosuperior}C_{0}(k)e^{-\frac{kz}{u}}}{u\left\langle k\right\rangle }-\dfrac{\gamma_{M}}{\beta_{M}^{0}u},
\end{equation}
and noticing that $i_{M}(0)\sim0$ implies $\theta_{M}(0)\sim0$, with
$z(0)=-\log(1-v_{M})$, one can integrate the above equation to obtain:
\begin{equation}
\begin{aligned}
\theta_{M} = \left(1-v_{M}\right)\left(1-\sum_{k}\dfrac{p_{k}}{\left\langle k\right\rangle }k\left(\left(1-v_{M}\right)e^{z}\right)^{-\frac{k}{u}}\right) -\dfrac{\gamma_{M}}{\beta_{M}^{0}u}\left(z+log(1-v_{M})\right).
\end{aligned}
\end{equation}
At the end of the measles outbreak $\theta_{M}(z(T))=0$, and one
finally obtains an equation for $s_{u}^{M}(T)$ that will obviously depend on the degree distribution $p_{k}$ of the network under consideration:
\begin{equation}
\begin{aligned}
0 = \left(1-v_{M}\right)\left(1-\sum_{k}\dfrac{p_{k}}{\left\langle k\right\rangle }k\left(\frac{1-v_{M}}{s_{u}^{M}(T)}\right)^{-\frac{k}{u}}\right) -\dfrac{\gamma_{M}}{\beta_{M}^{0}u}\log\left(\frac{1-v_{M}}{s_{u}^{M}(T)}\right).\label{Theta_z_General}
\end{aligned}
\end{equation}
Armed with Eqs.(\ref{Crit_General}), (\ref{sk_vs_sm}) and (\ref{Theta_z_General}),
we are now in a position to confidently explore the effects of different
networks topologies on the epidemic threshold.

\subsubsection*{B.1 Homogeneous networks}
In a homogeneous network all nodes have the same degree $\left\langle k\right\rangle $
(i.e., $p_{k}=\delta\left(k-\left\langle k\right\rangle \right)$),
so all connectivity classes are equal. Thus, we have $s_{k}^{M}=s_{M}$,
with $u=\left\langle k\right\rangle $, and Eq.(\ref{Theta_z_General})
is reduced to:
\begin{equation}
0 =\left(1-v_{M}\right)-s_{M}(T)-\dfrac{1}{R_{0}^{M}}\log\left(\frac{1-v_{M}}{s_{M}(T)}\right).
\end{equation}
Solving for $s_{M}(T)$ we find the following self-consistency equation:
\begin{equation}
s_{M}(T)=\left(1-v_{M}\right)e^{-R_{0}^{M}\left(\left(1-v_{M}\right)-s_{M}(T)\right)}\label{implicit_sm(T)}.
\end{equation}
Similarly, if all nodes have the same degree $\left\langle k\right\rangle $,
Eq.(\ref{Crit_General}) for the epidemic threshold can be easily reduced
to:
\begin{equation}
v_{M}^{\dagger\dagger}=\frac{v_{X}^{\dagger}}{v_{X}}-s_{M}(T)\label{crit_hom-1}
\end{equation}
Notice that, in the case in which $\mu\neq0$, the stationary-state
for the fraction of M-susceptibles is given by $s_{M}(T)=\frac{1}{R_{0}^{X}}$,
and we recover the epidemic threshold found in Eq.(\ref{crit_X_SI}). Eqs.(\ref{implicit_sm(T)}) and (\ref{crit_hom-1}) allows us to reproduce
an equivalent ---and in fact, very similar--- phase diagram to the one found for the demographic model
(see Fig.3 in the main text).

\subsubsection*{B.2 Erd\H{o}s-Rényi}
For an ER network and if $\left\langle k\right\rangle \ll N$, the binomial distribution typical of random networks can be approximated by a Poisson distribution, $p_{k}=\frac{\left\langle k\right\rangle ^{k}}{k!}e^{-\left\langle k\right\rangle }$.
Moreover, in the limit of very large networks, one can assume $u=1$ to be the minimum degree in the network. In terms of this distribution,
we can write: 
\begin{equation}
\begin{aligned}
\sum_{k} \dfrac{p_{k}}{\left\langle k\right\rangle }k\left(\left(1-v_{M}\right)e^{z}\right)^{-k} = e^{-\left\langle k\right\rangle }\sum_{k}\frac{\left\langle k\right\rangle ^{k-1}}{\left(k-1\right)!}\left(\left(1-v_{M}\right)e^{z}\right)^{-k}=\frac{e^{-\left(z+\left\langle k\right\rangle \right)}}{1-v_{M}}\exp\left(\frac{\left\langle k\right\rangle e^{-z}}{1-v_{M}}\right).
\end{aligned}
\end{equation}
Therefore, Eq.(\ref{Theta_z_General}) can be expressed as:
\begin{equation}
\begin{aligned}
0 = & s_{1}^{M}(T)e^{\left\langle k\right\rangle \left(\frac{s_{1}^{M}(T)}{1-v_{M}}-1\right)}-\frac{1}{R_{M}^{eff}}\log\left(\frac{s_{1}^{M}(T)}{1-v_{M}}\right)-\left(1-v_{M}\right),
\end{aligned}
\end{equation}
which has a trivial solution $s_{1}^{M}(T)=s_{k}^{M}(T)=1-v_{M}$,
representing the case in which no measles-epidemic has taken place.
On the other hand, making use of Eq.(\ref{sk_vs_sm}) we have:
\begin{equation}
\begin{aligned}
\left\langle k^{2}s_{M}(T)\right\rangle & =\left(1-v_{M}\right)e^{-\left\langle k\right\rangle}\sum_{k}\frac{k^{2}}{k!}\left(\frac{\left\langle k\right\rangle s_{1}^{M}(T)}{1-v_{M}}\right)^{k}\nonumber\\
 & = e^{-\left\langle k\right\rangle}\left\langle k\right\rangle s_{1}^{M}(T){_{1}}F{_{0}}\left(2,1;\frac{\left\langle k\right\rangle s_{1}^{M}(T)}{1-v_{M}}\right)\\
 & =\left\langle k\right\rangle s_{1}^{M}(T)\left(\frac{\left\langle k\right\rangle s_{1}^{M}(T)}{1-v_{M}}+1\right)e^{\left\langle k\right\rangle \left(\frac{s_{1}^{M}(\infty)}{1-v_{M}}-1\right)}.\nonumber
\end{aligned}
\end{equation}
where ${_{1}}F{_{0}}(a,b;z)$ is the generalized hypergeometric function \cite{abramowitz_handbook_1965}. Thus, using that $\left\langle k^{2}\right\rangle =\left\langle k\right\rangle \left(\left\langle k\right\rangle +1\right)$
for ER networks, the epidemic threshold condition is  written as:
\begin{equation}
\begin{aligned}
\dfrac{1}{R_{X}^{eff}} = \left(\left\langle k\right\rangle +1\right) \left(1-v_{M}^{\dagger\dagger}v_{X}\right) -v_{X}\left[s_{1}^{M}(T)\left(\frac{\left\langle k\right\rangle s_{1}^{M}(T)}{1-v_{M}^{\dagger\dagger}}+1\right)e^{\left\langle k\right\rangle \left(\frac{s_{1}^{M}(T)}{1-v_{M}^{\dagger\dagger}}-1\right)}\right],
\end{aligned}
\end{equation}
with $s_{1}^{M}(T)$ given by:
\begin{equation}
\begin{aligned}
s_{1}^{M}(T)e^{\left\langle k\right\rangle \left(\frac{s_{1}^{M}\left(T\right)}{1-v_{M}^{\dagger\dagger}}-1\right)}-\frac{1}{R_{M}^{eff}}\log\left(\frac{s_{1}^{M}(T)}{1-v_{M}^{\dagger\dagger}}\right)-\left(1-v_{M}^{\dagger\dagger}\right)=0.\label{s1m_equation-1}
\end{aligned}    
\end{equation}
In order to simplify a bit further these expressions, one can observe that
$s_{1}^{M}\left(T\right)$ must necessarily be a fraction
$q$ of the initial number of susceptibles (i.e., $s_{1}^{M}\left(T\right)=qs_{1}^{M}\left(0\right)=q\left(1-v_{M}\right)$) and then
rewrite the above equations as:
\begin{equation}
\begin{aligned}
\dfrac{1}{R_{X}^{eff}} = \left(\left\langle k\right\rangle +1\right)\left(1-v_{M}^{\dagger\dagger}v_{X}\right)-v_{X}e^{-\left\langle k\right\rangle }\left[q\left(1-v_{M}^{\dagger\dagger}\right)\left(\left\langle k\right\rangle q+1\right)e^{\left\langle k\right\rangle q}\right],
\end{aligned}    
\end{equation}
\begin{equation}
q\left(1-v_{M}^{\dagger\dagger}\right)e^{\left\langle k\right\rangle \left(q-1\right)}-\frac{1}{R_{M}^{eff}}\log\left(q\right)-\left(1-v_{M}^{\dagger\dagger}\right)=0.
\label{eq:Eq_1_ER_for_q}
\end{equation}
Regrouping terms in the epidemic threshold condition and defining $\zeta(q)=qv_{X}\left(1+\left\langle k\right\rangle q\right)e^{-\left\langle k\right\rangle \left(1-q\right)}$, we finally obtain:
\begin{equation}
v_{M}^{\dagger\dagger}(q)=\dfrac{1/R_{X}^{eff}-\left(1+\left\langle k\right\rangle \right)+\zeta(q)}{\zeta(q)-v_{X}\left(1+\left\langle k\right\rangle \right)}.    
\end{equation}
which can be inserted in Eq.(\ref{eq:Eq_1_ER_for_q}) to obtain a mathematical expression depending solely on $q$. We can then solve this equation to finally study the dependence of $v_M^{\dagger\dagger}$ with the network average degree, as done in the upper-right panel of Fig.6.

\subsubsection*{B.3. Power-law degree-distributed networks}
These networks are characterized by a probability distribution
of the form $p_{k}=C_{0}k^{-\alpha}$, where $C_{0}$ is determined
through the normalization condition $\sum_{k=u}^{k_{max}}p_{k}=1$. Furthermore, for networks with a large number of nodes, one can resort to the continuum formalism and replace the sums by integrals over $k$. Working in the limit of $N\rightarrow\infty$, we then have  $p_{k}=\left(\alpha-1\right)u^{\alpha-1}k^{-\alpha}\Theta(k-u)$.
In terms of this probability distribution function one can write:
\begin{equation}
\sum_{k} \dfrac{p_{k}}{\left\langle k\right\rangle }k\left(\left(1-v_{M}\right)e^{z}\right)^{-\frac{k}{u}}\approx \frac{\left(\alpha-1\right)u}{\left\langle k\right\rangle }\xi^{\alpha-2}\int_{\xi}^{\infty}e^{-x}x^{\left(2-\alpha\right)-1}dx=\left(\alpha-2\right)\xi^{\alpha-2}\Gamma\left(2-\alpha,\xi\right),
\end{equation}
where we have used $x=\frac{k}{u}\xi$, with $\xi=\log\left(\frac{1-v_{M}}{s_{u}^{M}}\right)$;
$\left\langle k\right\rangle =\left(\frac{1-\alpha}{2-\alpha}\right)u$ is the average degree of the network,
and $\Gamma(a,z)$ is the upper incomplete Gamma function \cite{abramowitz_handbook_1965}. Inserting
the above result into Eq.(\ref{Theta_z_General}) one obtains:
\begin{equation}
\dfrac{\xi}{uR_{M}^{eff}} = \left(1-v_{M}\right)\left(1-\left(\alpha-2\right)\xi^{\alpha-2}\Gamma\left(2-\alpha,\xi\right)\right).\label{condition_scfree}
\end{equation}
On the other hand, Eq.(\ref{sk_vs_sm})  can be used again to perform
an equivalent analysis of $\left\langle k^{2}s_{M}(T)\right\rangle $ to the one for ER networks:
\begin{equation}
\begin{aligned}
\left\langle k^{2}s_{M}(T)\right\rangle = & \left(1-v_{M}\right)\left(\alpha-1\right)u^{\alpha-1}\sum_{k=u}^{\infty}k^{2-\alpha}\left(\frac{s_{u}^{M}(T)}{1-v_{M}}\right)^{\frac{k}{u}}\\
 & \approx\left(1-v_{M}\right)\left(\alpha-1\right)u^{\alpha-1}\int_{u}^{\infty}k^{2-\alpha}e^{-\xi\frac{k}{u}}dk\\
 & =\left(1-v_{M}\right)\left(\alpha-1\right)u^{2}\xi^{\alpha-3}\int_{\xi}^{\infty}e^{-x}x^{\left(3-\alpha\right)-1}dx\\
 & =u^{2}\left(1-v_{M}\right)\left(\alpha-1\right)\xi^{\alpha-3}\Gamma\left(3-\alpha,\xi\right).
\end{aligned}    
\end{equation}
Finally, the vaccination threshold condition reduces to:
\begin{equation}
\dfrac{\left\langle k\right\rangle }{R_{X}^{eff}} = \left\langle k^{2}\right\rangle \left(1-v_{M}^{\dagger\dagger}v_{X}\right)-v_{X}u^{2}\left(1-v_{M}^{\dagger\dagger}\right)\left(\alpha-1\right)\xi^{\alpha-3}\Gamma\left(3-\alpha,\xi\right),\label{ep_thresh_SCFREE}
\end{equation}
which, using $\left\langle k^2 \right\rangle =\left(\frac{1-\alpha}{3-\alpha}\right)u^2$ and noticing that $v_{X}^{\dagger}=1-\frac{\left\langle k\right\rangle }{R_{X}^{eff}\left\langle k^{2}\right\rangle }$ is just the immunization threshold in the standard SIR model (derived in the thermodynamic limit) for power-law degree distributed networks \cite{pastor-satorras_immunization_2002}, can be written as:
\begin{equation}
v_{M}^{\dagger\dagger}=\dfrac{\left(\alpha-3\right)\xi^{\alpha-3}\Gamma(3-\alpha,\xi)-v_{X}^{\dagger}/v_{X}}{\left(\alpha-3\right)\xi^{\alpha-3}\Gamma(3-\alpha,\xi)-1}.\label{vm_dagdag_pwl}
\end{equation}
with a condition over $s_{u}^{M}(T)$ given in terms of $\xi$ by Eq.(\ref{condition_scfree}).
Although one could in principle solve the system given by Eqs.(\ref{condition_scfree}) and (\ref{ep_thresh_SCFREE}), we show in what follows that it is possible to derive the epidemic threshold from a slightly different approach that we believe is more informative. Beginning with the equations for the total fraction of susceptibles and resistants for X, and assuming as usual that measles outbreaks precede X outbreaks, we can write:
\begin{equation}
\begin{aligned}
\dfrac{ds_{X}^{k}}{dt} & =-\beta_{X}^{0}ks_{X}^{k}\theta_{X},\\
\dfrac{dr_{X}^{k}}{dt} & =\gamma_{X}i_{X}^{k}.
\end{aligned}
\end{equation}
Solving each equation between the time $T$ at which the X seed is
introduced and $t$:
\begin{equation}
\begin{aligned}
s_{X}^{k}(t)=s_{X}^{k}(T)e^{-\beta_{X}^{0}k\int_{T}^{t}\theta_{X}(t)dt},\\
r_{X}^{k}(t)=r_{X}^{k}(T)+\gamma_{X}\int_{T}^{t}i_{X}^{k}(t)dt.
\end{aligned}
\end{equation}
Defining $\phi_{X}=\int_{T}^{t}\theta_{X}(t)dt$ one can then write:
\begin{equation}
\begin{aligned}
\phi_{X} & =\sum_{k}\dfrac{p_{k}}{\left\langle k\right\rangle }k\int_{T}^{t}i_{X}^{k}(t)dt=\frac{1}{\left\langle k\right\rangle }\sum_{k}p_{k}k\left(\frac{r_{X}^{k}(t)-r_{X}^{k}(T)}{\gamma_{X}}\right).
\end{aligned}
\end{equation}
Using $i_{X}^{k}(t)=1-s_{X}^{k}(t)-r_{X}^{k}(t)$ and $s_{X}^{k}(T)=1-r_{X}^{k}(T)$:
\begin{equation}
\begin{aligned}
\dot{\phi}_{X} = & \dfrac{1}{\left\langle k\right\rangle }\sum_{k}p_{k}k i_{X}^{k}(t)=1-\dfrac{1}{\left\langle k\right\rangle }\sum_{k}p_{k}k\left(s_{X}^{k}(t)+r_{X}^{k}(t)\right)\\ 
= &-\gamma_{X}\phi+\dfrac{\left\langle s_{X}(T)k\right\rangle }{\left\langle k\right\rangle }-\dfrac{1}{\left\langle k\right\rangle }\sum_{k}p_{k}ks_{X}^{k}(T)e^{-\beta_{X}^{0}k\phi_{X}}.\nonumber
\end{aligned}
\end{equation}
One can now make use of Eq.(\ref{XandMsus}) and $\left\langle k^{2}\right\rangle =\left(\dfrac{1-\alpha}{3-\alpha}\right)u^{2}$, replacing again the infinite sums by integrals over $k$ to obtain:
\begin{equation}
\begin{aligned}
\dot{\phi}_{X} = & -\gamma_{X}\phi_{X}+1-v_{M}v_{X}+\\
 & -v_{X}\left(1-v_{M}\right)\left(\alpha-2\right)\xi^{\alpha-2}\Gamma\left(2-\alpha,\xi\right)\label{phi_dot_first}\\
 & -\left(1-v_{M}v_{X}\right)\left(\alpha-2\right)x^{\alpha-2}\Gamma(2-\alpha,x)\\
 & +v_{X}\left(1-v_{M}\right)\left(\alpha-2\right)\left(x+\xi\right)^{\alpha-2}\Gamma\left(2-\alpha,x+\xi\right).
\end{aligned}
\end{equation}
where for ease of notation we have defined $x=\beta_{X}^{0}u\phi_{X}$.
On the other hand, provided $s\neq0,-1,-2,...$; it is possible to
expand the upper incomplete gamma function as \cite{jameson_incomplete_2016}:
\begin{equation}
x^{\alpha-2}\Gamma\left(s,x\right) =x^{\alpha-2}\Gamma\left(s\right)-\sum_{n=0}^{\infty}\dfrac{\left(-1\right)^{n}x^{n}}{n!\left(s+n\right)}\label{expand_gamma},
\end{equation}
where $\Gamma(a)$ can be naturally extended to negative arguments
$a<0$ by mean of the recursive relation $\Gamma(a+1)=a\Gamma(a)$. Applying the former identity, and after several
manipulations, one can show that:
\begin{equation}
\begin{aligned}
\left(x+y\right)^{\alpha-2}\Gamma\left(s,x+y\right) = &x^{\alpha-2}\Gamma\left(s,x\right)+y^{\alpha-2}\Gamma\left(s,y\right)+\dfrac{1}{s}\\
 & +\Gamma(s)\left[\left(x+y\right)^{\alpha-2}-x^{\alpha-2}-y^{\alpha-2}\right]\\
 & +x\sum_{n=1}^{\infty}\dfrac{\left(-1\right)^{n}y^{n}}{n!\left(s+1+n\right)}-\frac{1}{2}x^{2}\sum_{n=1}^{\infty}\dfrac{\left(-1\right)^{n}y^{n}}{n!\left(s+2+n\right)}+O(x^{3}),
\end{aligned}
\end{equation}
which allow us to re-express Eq.(\ref{phi_dot_first}) as:
\begin{equation}
\begin{aligned}
\dot{\phi}_{X} = & -\gamma_{X}\phi_{X}+1-v_{X}\\
 & -\left(1-v_{X}\right)\left(\alpha-2\right)x^{\alpha-2}\Gamma(2-\alpha,x)\\
 & +v_{X}\left(1-v_{M}\right)\left(\alpha-2\right)x\sum_{n=1}^{\infty}\dfrac{\left(-1\right)^{n}\xi^{n}}{n!\left(3-\alpha+n\right)}\\
 & -\frac{1}{2}v_{X}\left(1-v_{M}\right)\left(\alpha-2\right)x^{2}\sum_{n=1}^{\infty}\dfrac{\left(-1\right)^{n}\xi^{n}}{n!\left(4-\alpha+n\right)}\\
 & +v_{X}\left(1-v_{M}\right)\left(\alpha-2\right)\Gamma(2-\alpha)\left[\left(x+\xi\right)^{\alpha-2}-x^{\alpha-2}-\xi^{\alpha-2}\right]+O(\phi_{X}^{3}).
\end{aligned}
\end{equation}
Expanding $\Gamma(2-\alpha,\beta_{X}^{0}u\phi_{X})$ and regrouping
everything we are finally left with:
\begin{equation}
\begin{aligned}
\dot{\phi}_{X} = & -\gamma_{X}\phi_{X}\\
 & -\left(1-v_{X}\right)\left(\alpha-2\right)\left[\dfrac{x}{\left(3-\alpha\right)}-\dfrac{x^{2}}{2\left(4-\alpha\right)}\right]\\
 & +v_{X}\left(1-v_{M}\right)\left(\alpha-2\right)x\left[\sum_{n=1}^{\infty}\dfrac{\left(-1\right)^{n}\xi^{n}}{n!\left(3-\alpha+n\right)}\right]\\
 & +\dfrac{1}{2}v_{X}\left(1-v_{M}\right)\left(\alpha-2\right)x^{2}\left[-\sum_{n=1}^{\infty}\dfrac{\left(-1\right)^{n}\xi^{n}}{n!\left(4-\alpha+n\right)}\right]\\
 & -\left(1-v_{X}v_{M}\right)\left(\alpha-2\right)\Gamma\left(2-\alpha\right)x^{\alpha-2}\\
 & +v_{X}\left(1-v_{M}\right)\left(\alpha-2\right)\Gamma(2-\alpha)\left[\left(x+\xi\right)^{\alpha-2}-\xi^{\alpha-2}\right]+O(\phi_{X}^{3}).
\end{aligned}
\end{equation}
The goal now is to rewrite the infinite series back in terms of Gamma functions using Eq.(\ref{expand_gamma}). To do so, one can add and subtract the missing $n=0$ terms that complete the series and use the identity $\Upsilon(s,x)=s\Gamma(s)-\Gamma(s,x)$ \cite{abramowitz_handbook_1965} to obtain:
\begin{equation}
\begin{aligned}
\dot{\phi}_{X}  = & -\gamma_{X}\phi_{X}-\left(1-v_{X}v_{M}\right)\left[x\dfrac{\left(\alpha-2\right)}{\left(3-\alpha\right)}-x^{2}\dfrac{\left(\alpha-2\right)}{2\left(4-\alpha\right)}\right]\\
 & +v_{X}\left(1-v_{M}\right)\left(\alpha-2\right)x\xi^{\alpha-3}\Upsilon\left(3-\alpha,\xi\right)\\
 & -\dfrac{1}{2}v_{X}\left(1-v_{M}\right)\left(\alpha-2\right)x^{2}\xi^{\alpha-4}\Upsilon\left(4-\alpha,\xi\right)\\
 & -\left(1-v_{X}v_{M}\right)\left(\alpha-2\right)\Gamma\left(2-\alpha\right)x^{\alpha-2}\\
 & +v_{X}\left(1-v_{M}\right)\left(\alpha-2\right)\Gamma(2-\alpha)\left[\left(x+\xi\right)^{\alpha-2}-\xi^{\alpha-2}\right]+O(\phi_{X}^{3}).
\end{aligned}
\end{equation}
We remark that, although the lower incomplete gamma function $\Upsilon(s,x)=\intop_{0}^{x}t^{s-1}e^{-t}dt$ is in principle defined for $Re(s)>0$, it is possible to extent its domain to
any non-integer $s<0$ by using $\Upsilon(s+1,x)=s\Upsilon(s,x)-x^{s}e^{-x}$
\cite{jameson_incomplete_2016}. Finally, although we have expanded $\dot{\phi}$ up to second order, the above expression can be easily generalized to any desired cutoff order $i_{max}$, writing it in a more compact way:
\begin{equation}
\begin{aligned}
\dot{\phi}_{X} = & -\gamma_{X}\phi_{X}+\left(\alpha-2\right)\sum_{i=2}^{i_{max}+1}\dfrac{\left(-1\right)^{i}x^{i-1}}{(i-1)!}\left[\dfrac{\left(1-v_{X}v_{M}\right)}{\left(\alpha-i-1\right)}+v_{X}\left(1-v_{M}\right)\xi^{\alpha-i-1}\Upsilon(i+1-\alpha,\xi)\right]\\
 & +\left(1-v_{M}v_{X}\right)x^{\alpha-2}\Gamma\left(3-\alpha\right)-v_{X}\left(1-v_{M}\right)\Gamma(3-\alpha)\left[\left(x+\xi\right)^{\alpha-2}-\xi^{\alpha-2}\right]+O(\phi_{X}^{i_{max}}).\label{eq:phiLong}
\end{aligned}
\end{equation}
Now, to obtain the epidemic threshold we analyze the stability of
the quiescent fixed point $\phi_{0}=0$, knowing that if $\dot{\phi}_{X}=f(\phi_{X})$, then $\lambda=\frac{\partial f}{\partial\phi}|_{\phi_{0}}=0$ gives
the bifurcation point. Before we proceed to study the different cases,
notice that in the vicinity of $\phi_{0}$ one can write the binomial
sum (for any real $\alpha$) as $\left(x+\xi\right)^{\alpha-2}=\xi^{\alpha-2}+\sum_{k=1}^{\infty}C\left(\alpha-2,k\right)x^{k}\xi^{\alpha-2-k}$, where $C\left(s,k\right)=\dfrac{s(s-1)(s-2)...(s+1-k)}{k!}$. This can simplify a bit Eq.(\ref{eq:phiLong}), which can be re-arranged ---after grouping together terms of the same order--- as: 

\begin{equation}
\begin{aligned}
\dot{\phi}_{X} = & -\gamma_{X}\phi_{X} + \left(\alpha-2\right)u\beta_{X}^{0}\left[\dfrac{\left(1-v_{X}v_{M}\right)}{\left(\alpha-3\right)}-v_{X}\left(1-v_{M}\right)\xi^{\alpha-3}\Gamma(3-\alpha,\xi)\right]\phi_{X}\\ 
& \left(\alpha-2\right)\sum_{i=3}^{i_{max}+1}
\dfrac{\left(-1\right)^{i}\left(u\beta_{X}^{0}\right)^{i-1}}{(i-1)!}\left[\dfrac{\left(1-v_{X}v_{M}\right)}{\left(\alpha-i-1\right)}-v_{X}\left(1-v_{M}\right)\xi^{\alpha-i-1}\Gamma(i+1-\alpha,\xi)\right]\phi_{X}^{i-1}\\
 & +\left(1-v_{M}v_{X}\right)\left(u\beta_{X}^{0}\phi_{X}\right)^{\alpha-2}\Gamma\left(3-\alpha\right)+O(\phi_{X}^{i_{max}}). \label{phi_complete}
\end{aligned}
\end{equation}
Two different cases need to be distinguished depending on whether the networks are scale-free or not:
\begin{enumerate}
\item If $2<\alpha<3$ (i.e. the network is scale-free) then the term $\phi_{X}^{\alpha-2}$ dominates with:
\begin{equation}
\lambda\sim\left(1-v_{M}v_{X}\right)\left(\alpha-2\right)x^{\alpha-3}\Gamma\left(3-\alpha\right),
\end{equation}
and since $\phi_{0}^{\alpha-3}\rightarrow\infty$, we will find $\lambda=0$ only if, trivially, $v_{X}=v_{M}=1$. This is the counterpart of saying that the epidemic threshold vanishes, in agreement with the well-known result in scale-free networks \cite{Barabasi:scfree,Pastor-Satorras}.

\item If otherwise $\alpha>3$, the leading order is given by the linear
term and we can write:

\begin{equation}
\lambda\sim \left(\alpha-2\right)u\beta_{X}^{0}\left[\dfrac{\left(1-v_{X}v_{M}\right)}{\left(\alpha-3\right)}-v_{X}\left(1-v_{M}\right)\xi^{\alpha-3}\Gamma(3-\alpha,\xi)\right] -\gamma_{X},
\end{equation}
One could, for instance, fix $v_{M}$ and write the epidemic threshold
in terms of $v_{X}$:
\begin{equation}
v_{X}^{\dagger\dagger}=\dfrac{v_{X}^{\dagger}}{v_{M}+\left(1-v_{M}\right)\left(\alpha-3\right)\xi^{\alpha-3}\Gamma(3-\alpha,\xi)}\label{vx_dagdag_pwl},
\end{equation}
where $v_{X}^{\dagger}=1-\frac{\left\langle k\right\rangle }{R_{X}^{eff}\left\langle k^{2}\right\rangle }=1-\frac{\left(\alpha-3\right)}{\left(uR_{X}^{eff}\right)\left(\alpha-2\right)}$
is the expected epidemic threshold for a SIR model with random vaccination
in a power-law degree-distributed network \cite{pastor-satorras_immunization_2002}. As a sanity check,
notice that one can retrieve the expected epidemic threshold for an independent disease in two different ways: by setting $v_{M}=1$, or by letting
$\xi\approx0$ and taking the asymptotic behavior of the upper-incomplete gamma function, $\xi^{\alpha-3}\Gamma(3-\alpha,\xi)\longrightarrow\left(\alpha-3\right)^{-1}$. It is also not difficult to show that (as expected) $v_{X}^{\dagger\dagger}\geq v_{X}^{SIR}$.
Using $\xi^{\alpha-3}\Gamma(3-\alpha,\xi)\leq\dfrac{e^{-\xi}}{\alpha-3}$
and $s_{M}^{u}(T)\leq1-v_{M}$ one can write:
\begin{equation}
\begin{aligned}
v_{M}&+\left(1-v_{M}\right)\left(\alpha-3\right)\xi^{\alpha-3}\Gamma(3-\alpha,\xi)\leq v_{M}+\left(1-v_{M}\right)e^{-\xi}=v_{M}+s_{M}^{u}(T)\leq1,
\end{aligned}
\end{equation}
which proves that the denominator in Eq.(\ref{vx_dagdag_pwl}) is always smaller or equal than one. On the other hand, it is also possible to derive the exact value of the
epidemic threshold in terms of the fraction of M-vaccinated individuals, recovering Eq.(\ref{vm_dagdag_pwl}). Remarkably, both $v_{X}^{\dagger\dagger}$ and $v_{M}^{\dagger\dagger}$
can be computed numerically using the relation between $v_{M}$ and $\xi$ given by Eq.(\ref{condition_scfree}), although in the main text the analysis is limited to the latter quantity.
\end{enumerate}

Finally, Eq.(\ref{phi_complete}) can help us to discern the nature of the bifurcation within our HMF approach. Developing the sum up to second order terms in $\phi_{X}$:
\begin{equation}
\begin{aligned}
\dot{\phi}_{X} = & -\gamma_{X}\phi_{X} + \left(\alpha-2\right)u\beta_{X}^{0}\left[\dfrac{\left(1-v_{X}v_{M}\right)}{\left(\alpha-3\right)}-v_{X}\left(1-v_{M}\right)\xi^{\alpha-3}\Gamma(3-\alpha,\xi)\right]\phi_{X}\\ 
& -\dfrac{1}{2}\left(\alpha-2\right)\left(u\beta_{X}^{0}\right)^{2}\left[\dfrac{\left(1-v_{X}v_{M}\right)}{\left(\alpha-4\right)}-v_{X}\left(1-v_{M}\right)\xi^{\alpha-4}\Gamma(4-\alpha,\xi)\right]\phi_{X}^{2}\\
 & +\left(1-v_{M}v_{X}\right)\left(u\beta_{X}^{0}\phi_{X}\right)^{\alpha-2}\Gamma\left(3-\alpha\right)+O(\phi_{X}^3). \label{phi_2nd_order}
\end{aligned}
\end{equation}

\begin{enumerate}
\item If $3<\alpha<4$: $\dot{\phi}\sim a\phi-\left|c\right|\phi^{\alpha-2}$
with $b\neq0$, which is the form of a transcritical bifurcation.
$\dot{\phi}=0$ admits therefore a solution with arbitrary small $\phi\neq0$
when we are slightly above the epidemic threshold, and the transition
is continuous.
\item If $4<\alpha$: $\dot{\phi}\sim a\phi+b\phi^{2}$. Using that  $\xi^{\alpha-4}\Gamma(3-\alpha,\xi)\leq\dfrac{e^{-\xi}}{\alpha-4}$ \cite{jameson_incomplete_2016} and $s_{M}^{u}(T)\leq1-v_{M}$ one can write: 
\begin{equation}
\dfrac{\left(1-v_{X}v_{M}\right)}{\left(\alpha-4\right)}-v_{X}\left(1-v_{M}\right)\xi^{\alpha-4}\Gamma(4-\alpha,\xi)  \geq \dfrac{1-v_{X}v_{M}-v_{X}s_{M}^{u}(T)}{\left(\alpha-4\right)}
\geq  \dfrac{1-v_{X}}{\left(\alpha-4\right)}\geq0
\end{equation}
It follows that $b<0$ and the transition is therefore continuous.
\end{enumerate}



\bibliographystyle{ieeetr}  
\bibliography{references}  

\begin{thebibliography}{10}

\bibitem{noauthor_who_nodate}
``{WHO} {\textbar} {Global} {Vaccine} {Action} {Plan} 2011-2020,'' 2012.
\newblock Publisher: World Health Organization.

\bibitem{MV}
D.~Griffin, ``Immune responses during measles virus infection,'' in {\em
  Measles virus}, pp.~117--134, Springer, 1995.

\bibitem{noauthor_who_nodate-1}
``{WHO} {\textbar} {Measles} and {Rubella} {Surveillance} {Data},'' 2020.
\newblock Publisher: World Health Organization.

\bibitem{Eva}
E.~Frederick, ``How measles causes the body to ‘forget’past infections,''
  2019.

\bibitem{Italy}
A.~Siani, ``Measles outbreaks in italy: A paradigm of the re-emergence of
  vaccine-preventable diseases in developed countries,'' {\em Preventive
  medicine}, vol.~121, pp.~99--104, 2019.

\bibitem{noauthor_who_nodate-2}
``{WHO} {\textbar} {More} than 117 million children at risk of missing out on
  measles vaccines, as {COVID}-19 surges,'' 2020.
\newblock Publisher: World Health Organization.

\bibitem{thompson_evolution_2016}
K.~M. Thompson, ``Evolution and {Use} of {Dynamic} {Transmission} {Models} for
  {Measles} and {Rubella} {Risk} and {Policy} {Analysis},'' {\em Risk
  Analysis}, vol.~36, pp.~1383--1403, July 2016.
\newblock Publisher: John Wiley \& Sons, Ltd.

\bibitem{de_vries_measles_2012}
R.~D. de~Vries, S.~McQuaid, G.~van Amerongen, S.~Yüksel, R.~J. Verburgh, A.~D.
  M.~E. Osterhaus, W.~P. Duprex, and R.~L. de~Swart, ``Measles immune
  suppression: lessons from the macaque model,'' {\em PLoS pathogens}, vol.~8,
  no.~8, p.~e1002885, 2012.

\bibitem{de_vries_measles_2014}
R.~D. de~Vries and R.~L. de~Swart, ``Measles immune suppression: functional
  impairment or numbers game?,'' {\em PLoS pathogens}, vol.~10, p.~e1004482,
  Dec. 2014.

\bibitem{mina_long-term_2015}
M.~J. Mina, C.~J.~E. Metcalf, R.~L. de~Swart, A.~D. M.~E. Osterhaus, and B.~T.
  Grenfell, ``Long-term measles-induced immunomodulation increases overall
  childhood infectious disease mortality,'' {\em Science (New York, N.Y.)},
  vol.~348, pp.~694--699, May 2015.

\bibitem{mina_measles_2019}
M.~J. Mina, T.~Kula, Y.~Leng, M.~Li, R.~D.~d. Vries, M.~Knip, H.~Siljander,
  M.~Rewers, D.~F. Choy, M.~S. Wilson, H.~B. Larman, A.~N. Nelson, D.~E.
  Griffin, R.~L.~d. Swart, and S.~J. Elledge, ``Measles virus infection
  diminishes preexisting antibodies that offer protection from other
  pathogens,'' {\em Science}, vol.~366, pp.~599--606, Nov. 2019.

\bibitem{petrova_incomplete_2019}
V.~N. Petrova, B.~Sawatsky, A.~X. Han, B.~M. Laksono, L.~Walz, E.~Parker,
  K.~Pieper, C.~A. Anderson, R.~D.~d. Vries, A.~Lanzavecchia, P.~Kellam, V.~v.
  Messling, R.~L.~d. Swart, and C.~A. Russell, ``Incomplete genetic
  reconstitution of {B} cell pools contributes to prolonged immunosuppression
  after measles,'' {\em Science Immunology}, vol.~4, Nov. 2019.
\newblock Publisher: Science Immunology Section: Research Article.

\bibitem{IA-Netherlands}
B.~M. Laksono, R.~D. de~Vries, R.~J. Verburgh, E.~G. Visser, A.~de~Jong, P.~L.
  Fraaij, W.~L. Ruijs, D.~F. Nieuwenhuijse, H.-J. van~den Ham, M.~P. Koopmans,
  {\em et~al.}, ``Studies into the mechanism of measles-associated immune
  suppression during a measles outbreak in the netherlands,'' {\em Nature
  communications}, vol.~9, no.~1, pp.~1--10, 2018.

\bibitem{miller_frequency_1964}
D.~L. Miller, ``Frequency of {Complications} of {Measles}, 1963,'' {\em Br Med
  J}, vol.~2, pp.~75--78, July 1964.
\newblock Publisher: British Medical Journal Publishing Group Section: Papers
  and Originals.

\bibitem{beckford_factors_1985}
A.~P. Beckford, R.~O. Kaschula, and C.~Stephen, ``Factors associated with fatal
  cases of measles. {A} retrospective autopsy study,'' {\em South African
  Medical Journal = Suid-Afrikaanse Tydskrif Vir Geneeskunde}, vol.~68,
  pp.~858--863, Dec. 1985.

\bibitem{anderson_infectious_1992}
R.~M. Anderson and R.~M. May, {\em Infectious {Diseases} of {Humans}:
  {Dynamics} and {Control}}.
\newblock Oxford, New York: Oxford University Press, Aug. 1992.

\bibitem{Hethcote}
H.~W. Hethcote, ``The mathematics of infectious diseases,'' {\em SIAM review},
  vol.~42, no.~4, pp.~599--653, 2000.

\bibitem{Anderson-book}
R.~M. Anderson, B.~Anderson, and R.~M. May, {\em Infectious diseases of humans:
  dynamics and control}.
\newblock Oxford university press, 1992.

\bibitem{Murray-book}
J.~D. Murray, {\em Mathematical biology: I. An introduction}, vol.~17.
\newblock Springer Science \& Business Media, 2007.

\bibitem{Pastor-Satorras}
R.~Pastor-Satorras, C.~Castellano, P.~Van~Mieghem, and A.~Vespignani,
  ``Epidemic processes in complex networks,'' {\em Reviews of modern physics},
  vol.~87, no.~3, p.~925, 2015.

\bibitem{Gillespie}
D.~T. Gillespie, ``Exact stochastic simulation of coupled chemical reactions,''
  {\em The journal of physical chemistry}, vol.~81, no.~25, pp.~2340--2361,
  1977.

\bibitem{fontanet_covid-19_2020}
A.~Fontanet and S.~Cauchemez, ``{COVID}-19 herd immunity: where are we?,'' {\em
  Nature Reviews Immunology}, vol.~20, pp.~583--584, Oct. 2020.
\newblock Number: 10Publisher: Nature Publishing Group.

\bibitem{bauch_transients_2003}
C.~T. Bauch and D.~J.~D. Earn, ``Transients and attractors in epidemics,'' {\em
  Proceedings of the Royal Society of London. Series B: Biological Sciences},
  vol.~270, pp.~1573--1578, Aug. 2003.
\newblock Publisher: Royal Society.

\bibitem{guerra_basic_2017}
F.~M. Guerra, S.~Bolotin, G.~Lim, J.~Heffernan, S.~L. Deeks, Y.~Li, and N.~S.
  Crowcroft, ``The basic reproduction number ({R} 0 ) of measles: a systematic
  review,'' {\em The Lancet Infectious Diseases}, vol.~17, pp.~e420--e428, Dec.
  2017.

\bibitem{li_substantial_2020}
R.~Li, S.~Pei, B.~Chen, Y.~Song, T.~Zhang, W.~Yang, and J.~Shaman,
  ``Substantial undocumented infection facilitates the rapid dissemination of
  novel coronavirus ({SARS}-{CoV}-2),'' {\em Science (New York, N.y.)},
  vol.~368, pp.~489--493, May 2020.

\bibitem{EpidemicsRandom_Mollison}
A.~Barbour and D.~Mollison, ``Epidemics and random graphs,'' in {\em Stochastic
  Processes in Epidemic Theory} (J.-P. Gabriel, C.~Lef{\`e}vre, and P.~Picard,
  eds.), (Berlin, Heidelberg), pp.~86--89, Springer, 1990.

\bibitem{Barabasi:scfree}
A.-L. Barab{\'a}si and R.~Albert, ``Emergence of scaling in random networks,''
  {\em Science}, vol.~286, no.~5439, pp.~509--512, 1999.

\bibitem{newman_spread_2002}
M.~E.~J. Newman, ``The spread of epidemic disease on networks,'' {\em Physical
  Review E}, vol.~66, p.~016128, July 2002.

\bibitem{SHIRLEY2005287}
M.~D. Shirley and S.~P. Rushton, ``The impacts of network topology on disease
  spread,'' {\em Ecological Complexity}, vol.~2, no.~3, pp.~287 -- 299, 2005.

\bibitem{keeling_networks_2005}
M.~J. Keeling and K.~T. Eames, ``Networks and epidemic models,'' {\em Journal
  of The Royal Society Interface}, vol.~2, pp.~295--307, Sept. 2005.

\bibitem{barrat_dynamical_2008}
A.~Barrat, M.~Barthélemy, and A.~Vespignani, {\em Dynamical {Processes} on
  {Complex} {Networks}}.
\newblock Cambridge: Cambridge University Press, 2008.

\bibitem{pastor-satorras_epidemic_2001}
R.~Pastor-Satorras and A.~Vespignani, ``Epidemic {Spreading} in {Scale}-{Free}
  {Networks},'' {\em Physical Review Letters}, vol.~86, pp.~3200--3203, Apr.
  2001.

\bibitem{Erdos:1959:pmd}
P.~Erd\"os and A.~R\'enyi, ``On random graphs i,'' {\em Publicationes
  Mathematicae Debrecen}, vol.~6, pp.~290--297, 1959.

\bibitem{newman_structure_2003}
M.~E.~J. Newman, ``The {Structure} and {Function} of {Complex} {Networks},''
  {\em SIAM Review}, vol.~45, pp.~167--256, Jan. 2003.

\bibitem{newman_networks_2018}
M.~Newman, {\em Networks}.
\newblock Oxford University Press, July 2018.
\newblock Google-Books-ID: YdZjDwAAQBAJ.

\bibitem{kitsak_identification_2010}
M.~Kitsak, L.~K. Gallos, S.~Havlin, F.~Liljeros, L.~Muchnik, H.~E. Stanley, and
  H.~A. Makse, ``Identification of influential spreaders in complex networks,''
  {\em Nature Physics}, vol.~6, pp.~888--893, Nov. 2010.

\bibitem{madar_immunization_2004}
N.~Madar, T.~Kalisky, R.~Cohen, D.~ben Avraham, and S.~Havlin, ``Immunization
  and epidemic dynamics in complex networks,'' {\em The European Physical
  Journal B}, vol.~38, pp.~269--276, Mar. 2004.

\bibitem{cohen_efficient_2003}
R.~Cohen, S.~Havlin, and D.~Ben-Avraham, ``Efficient immunization strategies
  for computer networks and populations,'' {\em Physical Review Letters},
  vol.~91, p.~247901, Dec. 2003.

\bibitem{pastor-satorras_immunization_2002}
R.~Pastor-Satorras and A.~Vespignani, ``Immunization of complex networks,''
  {\em Physical Review E}, vol.~65, p.~036104, Feb. 2002.

\bibitem{grassberger_phase_2016}
P.~Grassberger, L.~Chen, F.~Ghanbarnejad, and W.~Cai, ``Phase transitions in
  cooperative coinfections: {Simulation} results for networks and lattices,''
  {\em Physical Review E}, vol.~93, p.~042316, Apr. 2016.

\bibitem{cui_mutually_2017}
P.-B. Cui, F.~Colaiori, and C.~Castellano, ``Mutually cooperative epidemics on
  power-law networks,'' {\em Physical Review E}, vol.~96, p.~022301, Aug. 2017.

\bibitem{castellano_thresholds_2010}
C.~Castellano and R.~Pastor-Satorras, ``Thresholds for {Epidemic} {Spreading}
  in {Networks},'' {\em Physical Review Letters}, vol.~105, p.~218701, Nov.
  2010.

\bibitem{barabasi2016network}
A.-L. Barabási and M.~Pósfai, {\em Network science}.
\newblock Cambridge: Cambridge University Press, 2016.

\bibitem{may_infection_2001}
R.~M. May and A.~L. Lloyd, ``Infection dynamics on scale-free networks,'' {\em
  Physical Review. E, Statistical, Nonlinear, and Soft Matter Physics},
  vol.~64, p.~066112, Dec. 2001.

\bibitem{peng_vaccination_2013}
X.-L. Peng, X.-J. Xu, X.~Fu, and T.~Zhou, ``Vaccination intervention on
  epidemic dynamics in networks,'' {\em Physical Review E}, vol.~87, p.~022813,
  Feb. 2013.

\bibitem{van_herwaarden_stochastic_1997}
O.~A. van Herwaarden, ``Stochastic epidemics: the probability of extinction of
  an infectious disease at the end of a major outbreak,'' {\em Journal of
  Mathematical Biology}, vol.~35, pp.~793--813, Aug. 1997.

\bibitem{meerson_wkb_2009}
B.~Meerson and P.~V. Sasorov, ``{WKB} theory of epidemic fade-out in stochastic
  populations,'' {\em Physical Review E}, vol.~80, p.~041130, Oct. 2009.
\newblock Publisher: American Physical Society.

\bibitem{hartfield_introducing_2013}
M.~Hartfield and S.~Alizon, ``Introducing the {Outbreak} {Threshold} in
  {Epidemiology},'' {\em PLoS Pathogens}, vol.~9, June 2013.

\bibitem{chen_outbreaks_2013}
L.~Chen, F.~Ghanbarnejad, W.~Cai, and P.~Grassberger, ``Outbreaks of
  coinfections: {The} critical role of cooperativity,'' {\em EPL (Europhysics
  Letters)}, vol.~104, p.~50001, Dec. 2013.

\bibitem{cai_avalanche_2015}
W.~Cai, L.~Chen, F.~Ghanbarnejad, and P.~Grassberger, ``Avalanche outbreaks
  emerging in cooperative contagions,'' {\em Nature Physics}, vol.~11,
  pp.~936--940, Nov. 2015.

\bibitem{janssen_first-order_2016}
H.-K. Janssen and O.~Stenull, ``First-order phase transitions in outbreaks of
  co-infectious diseases and the extended general epidemic process,'' {\em EPL
  (Europhysics Letters)}, vol.~113, p.~26005, Jan. 2016.
\newblock arXiv: 1602.01786.

\bibitem{catastrophic1}
M.~Scheffer, S.~Carpenter, J.~A. Foley, C.~Folke, and B.~Walker, ``Catastrophic
  shifts in ecosystems,'' {\em Nature}, vol.~413, no.~6856, pp.~591--596, 2001.

\bibitem{catastrophic2}
P.~V. Mart{\'\i}n, J.~A. Bonachela, S.~A. Levin, and M.~A. Mu{\~n}oz, ``Eluding
  catastrophic shifts,'' {\em Proceedings of the National Academy of Sciences},
  vol.~112, no.~15, pp.~E1828--E1836, 2015.

\bibitem{sanz_dynamics_2014}
J.~Sanz, C.-Y. Xia, S.~Meloni, and Y.~Moreno, ``Dynamics of {Interacting}
  {Diseases},'' {\em Physical Review X}, vol.~4, p.~041005, Oct. 2014.

\bibitem{dezso_halting_2002}
Z.~Dezső and A.-L. Barabási, ``Halting viruses in scale-free networks,'' {\em
  Physical Review E}, vol.~65, p.~055103, May 2002.

\bibitem{strogatz_nonlinear_2000}
S.~H. Strogatz, {\em Nonlinear {Dynamics} and {Chaos}: {With} {Applications} to
  {Physics}, {Biology}, {Chemistry} and {Engineering}}.
\newblock Westview, 2000.

\bibitem{Bogacki_Rhunge}
P.~Bogacki and L.~Shampine, ``A 3(2) pair of runge - kutta formulas,'' {\em
  Applied Mathematics Letters}, vol.~2, no.~4, pp.~321 -- 325, 1989.

\bibitem{kiss_mathematics_2017}
I.~Z. Kiss, J.~Miller, and P.~L. Simon, {\em Mathematics of {Epidemics} on
  {Networks}: {From} {Exact} to {Approximate} {Models}}.
\newblock Interdisciplinary {Applied} {Mathematics}, Springer International
  Publishing, 2017.

\bibitem{bornholdt_handbook_2006}
S.~Bornholdt and H.~G. Schuster, {\em Handbook of {Graphs} and {Networks}:
  {From} the {Genome} to the {Internet}}.
\newblock John Wiley \& Sons, Mar. 2006.
\newblock Google-Books-ID: kZo5ZbzkXOIC.

\bibitem{catanzaro_generation_2005}
M.~Catanzaro, M.~Boguñá, and R.~Pastor-Satorras, ``Generation of uncorrelated
  random scale-free networks,'' {\em Physical Review E}, vol.~71, p.~027103,
  Feb. 2005.

\bibitem{lucas_exact_2012}
A.~Lucas, ``Exact mean field dynamics for epidemic-like processes on
  heterogeneous networks,'' {\em arXiv:1206.6294 [physics]}, June 2012.
\newblock arXiv: 1206.6294.

\bibitem{abramowitz_handbook_1965}
M.~Abramowitz and I.~A. Stegun, {\em Handbook of {Mathematical} {Functions}:
  {With} {Formulas}, {Graphs}, and {Mathematical} {Tables}}.
\newblock Courier Corporation, Jan. 1965.
\newblock Google-Books-ID: MtU8uP7XMvoC.

\bibitem{jameson_incomplete_2016}
G.~J.~O. Jameson, ``The incomplete gamma functions,'' {\em The Mathematical
  Gazette}, vol.~100, pp.~298--306, July 2016.

\end{thebibliography}

\end{document}